\documentclass[sigconf,nonacm]{acmart} % preprint
\AtBeginDocument{%
  }

\setcopyright{acmlicensed}
\copyrightyear{2026}
\acmYear{2026}
\acmDOI{XXXXXXX.XXXXXXX}
\usepackage{multirow}
\usepackage{subcaption}
\def\thickhline{\noalign{\hrule height1pt}}
\usepackage{boldline}
\usepackage{graphicx}
\usepackage{adjustbox}

\usepackage{soul}
\usepackage[dvipsnames]{xcolor}
\definecolor{alizarin}{rgb}{0.82, 0.1, 0.26}
\begin{document}

\title{Priming, Path-dependence, and Plasticity: Understanding the molding of user-LLM interaction and its implications from (many) chat logs in the wild}


\author{Shengqi Zhu}
\email{sz595@cornell.edu}
\orcid{0009-0002-5660-3241}
\affiliation{%
 \institution{Cornell University}
 \city{Ithaca, NY}
 \country{USA}}
 
\author{Jeffrey M. Rzeszotarski}
\email{jeff.rzeszotarski@gmail.com}
\orcid{0000-0002-4317-9501}
\affiliation{%
 \institution{Loyola University Maryland}
 \city{Baltimore, MD}
 \country{USA}}

\author{David Mimno}
\email{mimno@cornell.edu}
\orcid{0000-0001-7510-9404}
\affiliation{%
 \institution{Cornell University}
 \city{Ithaca, NY}
 \country{USA}}



\begin{abstract}
User interactions with LLMs are shaped by prior experiences and individual exploration, but in-lab studies do not provide system designers with visibility into these in-the-wild factors.
This work explores a new approach to studying real-world user-LLM interactions through large-scale chat logs from the wild.
Through analysis of 140K chatbot sessions from 7,955 anonymized global users over time, 
we demonstrate key patterns in user expressions despite varied tasks:
(1) LLM users are not \textit{tabula rasa}, nor are they constantly adapting; rather, interaction patterns form and stabilize rapidly through individual early trajectories;
(2) Longitudinal outcomes, such as recurring text patterns and retention rates, are strongly correlated with early exploration;
(3) Parallel dynamics are present, including organizing expressions by task types such as emotional support, or in response to model-version updates.
These results present an ``agency paradox'': despite LLM input spaces being unconstrained and user-driven, we in fact see less user exploration.
We call for design consideration surrounding the molding procedure and its incorporation in future research.

\end{abstract}

%TC:ignore




\received{2026}
% \received[revised]{}
% \received[accepted]{}

%TC:endignore

\maketitle

\section{Introduction}

The appeal of chatbot-style language model interaction often lies in its flexibility and open-ended nature.
Real-world user-LLM interactions are free-form, user-driven, and longitudinal~\cite{zhaowildchat,tamkin2024clio,handa2025economictasksperformedai}. Users may (or may not) freely explore the input space, iterate on phrasing, and more importantly, develop \textit{expression} patterns that may be reused over time and across tasks~\cite{zhu2025requestmaking,schneider2025mental}. 
There are many potential reasons why interactions in the real world may appear ``suboptimal'': users can be hurried, curious, playful, or simply unmotivated~\cite{zhaowildchat,zhenglmsys}.
In such cases, one may yield to unconscious and long-term factors such as habitual language use or prior experience.
This very form of interaction \textit{in the wild} plays a vital role in shaping one's perception and mental model that are carried into any subsequent interaction, including lab experiments.

Nonetheless, such ``background'' procedures remains largely unexplored. Human-AI Interaction studies in the LLM context often operate on more structured and ``ideal'' sessions, with users instructed to actively interact with a controlled system and prompted for task-oriented feedback~\cite{jiang2024human}. Current research has yet to propose systematic models for how users familiarize themselves and develop interaction strategies \textit{for their own purposes} and \textit{on their own terms}. Less is known about shared patterns (or lack thereof) \textit{at scale}: across large numbers of users, underneath distinct request contents, and over an extended period of time. For instance, ``prior experience'' is often modeled (if at all) as a simple categorical condition (e.g., ``novice vs. expert'') rather than as a complex factor that shapes the basis of users' perception, repertoire, and decision-making.

In this work, we take the first steps to understand the procedures of user-LLM conversations in noisy real-world scenarios, especially the underlying fundamentals shared across varied tasks.\footnote{The code base for our paper will be open-sourced upon acceptance.}
Unlike conventional researcher-designed experiments/systems, we explore a novel approach centering users and the data they naturally created: examining large-scale user-LLM chat logs ($\sim$140K dialogs from $\sim$8K users) from the wild with text-based analysis.
More specifically, we quantitatively model the language patterns used to deliver requests, the most direct indicator of open, in-the-wild interaction modes.
We particularly highlight the early \textit{priming} and rapid \textit{molding} of users' explorations, demonstrating how a user's initial interactions start a cascade of behaviors that fundamentally influence their long-term usage patterns and life cycle.
While current data resources cannot assess causal claims, we are able to make substantial observations about patterns of user behavior that are only possible in this fashion of large-scale real-world user analysis.

We begin by processing the recorded logs to separate \textit{expression} and \textit{task/request content}, following the data and protocols provided by \citet{zhu2025requestmaking}'s ReCCRE, a variation of the million-scale WildChat dataset~\cite{zhaowildchat} that features a clean division between task-relevant and task-independent components.
We confirm the \textit{convergence} of all components as a user gains more experience. 
Contrary to the common assumption of an actively and constantly adapting user, user-LLM interactions in the wild appear surprisingly ``sticky'' and are molded by the (more random) attempts early on.
However, \textit{what} the users are doing (the requests) and \textit{how} they are doing them (the expressions) show distinct patterns.
In general, users pick up patterns within a handful of dialog sessions, and use a rather limited set of expressions to accommodate the much more diverse space of tasks.
Further, we uncover important potential influences of \textit{early interactions} given the rapid molding of user behaviors.
Text patterns used early on reoccur significantly (5--50$\times$) more often, and more diverse expressions in early interaction are directly associated with longer user lifespans.
More studies are required to incorporate complex dynamics that could alter the user pattern once stabilized, as we show two examples of parallel dynamics and perturbation later on:
(1) Users simultaneously organize different expressions for different task types, most significantly for Social and Emotional Support;
(2) Over time, users vary their expressions in response to model updates, as switching from GPT-3.5 to GPT-4 triggers a return to exploratory behaviors.

Our work contributes both methodological advances and empirical findings.
First, we systematically validate a new user-LLM research method to draw user-centered insights from large-scale user-LLM chat logs~\cite{zhu2025data,zhaowildchat,zhu2025requestmaking}, extending recent data and pre-processing frameworks (\S\ref{subsec:dataset_basics}) and applying practical text-analysis setups (\S\ref{subsec:experimental_setup}).
With this methodology, we systematically model the early molding of user expressions (\S\ref{subsec:molding_user_input}), delineate the profound effects of earliest interactions (\S\ref{subsec:Experiments_early_interaction}), and discuss other key dynamics and possible interventions in the molding process (\S\ref{subsec:Experiments_interventions}).
We encourage interaction designers to recognize that users come to new systems with strong and sticky expectations about how to format their prompts.
LLM interaction design should actively engage with the momentum of early interactions (\S\ref{subsec:Discussion_design_with_early_interaction}), as users may not be aware of alternative paths or their impacts.
Designers should also be cautious in adopting a common assumption of an ``actively learning user'' (\S\ref{subsec:Discussion_illusion_active_user}).
Finally, we propose to consider HAI research in the presence of these underexplored but ubiquitous priors --- as an important factor working with real-world users, a new objective towards comprehensive user experience, as well as a deeply rooted factor impacting researchers themselves (\S\ref{subsec:Discussion_future_research_with_priors}).

\section{Background \& Related Work}

We begin by overviewing the crucial interdisciplinary backgrounds of user behavioral adaptation in open-ended interface (\S\ref{subsec:RelatedWork_adaptation_open-ended}) and priming effects in such setups (\S\ref{subsec:RelatedWork_early_impacts_non-optimality}).
From there, we discuss the specific setups of our work enabled by recent data sources: from-the-wild data, large-scale user-LLM chat logs, and task-independency.

\subsection{Adaptation and Exploration in Open-Ended Interfaces}
\label{subsec:RelatedWork_adaptation_open-ended}
Interactions with natural language interfaces (NLIs) differ from traditional graphical UIs, where
a user internalize menu structures, build muscle memory for shortcuts, and learn the visible ``rules of the game''~\cite{card1983,norman2013everday}.
This is often learned by \textit{direct manipulation}~\cite{Shneiderman1983DirectMA} with incremental visible objects and reversible actions, creating a sense of control and mastery.
In contrast, conversational systems remove the visible scaffolding and instead present users with an unbounded expression space.
Users of conversational agents~\cite{luger2016like} and smart speakers~\cite{porcheron2018voice} struggle to discover system capabilities and form accurate mental models without persistent visual cues.
The emergence of Generative AI and LLMs with user-driven \textit{prompts} as input has amplified these challenges across multiple dimensions.
Users, especially non-experts, struggle with fundamental prompt formulation and explore more opportunistically rather than systematically~\cite{zp2023whyjohnny,arawjo2024chainforge}.
They often find it difficult to arrive at their desired outcome through iterations of prompts in downstream tasks such as data labeling~\cite{he2025promptingdark} and UX design~\cite{zp2023herdingAIcat}.
Faced with imposed megacognitive demands~\cite{tankelevitch2024megacognitive}, users can develop false understandings of system capabilities and how LLMs process information~\cite{zhang2024fairgame,liao2024ai}, and these inaccurate beliefs may be prolonged~\cite{gero2024sensemaking}.

\subsection{Early Impacts, Path Dependence, and Non-Optimality}
\label{subsec:RelatedWork_early_impacts_non-optimality}
Users' discovery processes are rarely a systematic optimization; instead, they are guided by individual cognitive principles of efficiency and habits that may yield idiosyncratic, non-optimal situations.
\citet{simon1956rational} models users as \textit{satisficers} seeking ``good enough'' (rather than ``the best'') solutions within bounded rationality.
\citet{carroll1987paradox} describes the \textit{paradox of the active user} where users' motivation to act before studying a system leads to error-prone and incomplete explorations.
Research on skill acquisition describes different ``phases'' and reports the disproportionate influence of early decisions on later performance~\cite{anderson1982acquisition,fitts1967human}.
These findings align with the \textit{path dependence} theory: originally explaining technological lock-ins like QWERTY keyboards~\cite{david1985clio}, path dependence describes how minor early choices lead to self-reinforcing feedback loops with major influences on outcomes~\cite{arthur1989competing,luchins1942mechanization}.
However, similar models and implications for individual user behavior in open-ended interfaces, and especially the versatile NLIs, remain underexplored.

\

Summing up \S\ref{subsec:RelatedWork_adaptation_open-ended} and \S\ref{subsec:RelatedWork_early_impacts_non-optimality}, our work focuses on the complexity at the overlap of these theoretical foundations: while HCI research seeks sufficient user control~\cite{Shneiderman1983DirectMA,nielsen1994usability}, excessive flexibility may on the other hand overwhelm users.
This is sometimes described as a ``paradox of choice'', originating from a consumer model in economics: too many options may actually lead to decision paralysis and reduced satisfaction~\cite{schwartz2015paradox}.
We examine such influences in the LLM context, which allows even more (in fact, infinite) options with the flexibility of natural language input.
Without visible affordances, users need to construct their own interaction frameworks, which we show associate heavily with a limited number of early attempts.
We describe this as an \textit{agency paradox}: an interface designed with maximum expressive freedom may, in practice, yield users with reduced motivation to explore, and eventually leads to less diverse interaction modes.

\subsection{Novel Setups in This Work: Contexts, Motivations, and Methodologies}
Our work is built upon an emerging data paradigm~\cite{zhu2025data}: mining user behavior patterns from large-scale natural interactions with LLMs.
We discuss the key contexts and implications of this new fashion, before providing more technical details in \S\ref{sec:dataset_and_setup}.
We start from the necessity of incorporating prior experience, underscore the (over)simplification of users' prior experience in typical experiment setups, and introduce recent threads of research modeling Human-AI Interaction as a \textit{longitudinal} process.
From there, we highlight the complexities underlying real-world user-LLM interactions and introduce the status quo of leveraging user data \textit{from the wild}.% and the available resources from Natural Language Processing (NLP).

\subsubsection{Incorporation of Prior Experience in Human-AI Interaction}
A critical challenge in HAI research is to consider the varied ``prior experience'' of users in meaningful ways.
Current approaches typically consider self-reported categorical or binary classifications from screening questionnaires, such as by duration (``months of use'')~\cite{jo2024carecall,yu2024parent}, frequency (``daily user'')~\cite{laban2024editing,zhang2025navfog}, expertise levels (``novice vs. experts'')~\cite{chen2024novice,mahdavi2024image}, or familiarity (``familiar/not familiar'', ``no experience/has experience'')~\cite{belghith2024schooler,zheng2025library,fu2024self}.
While this paradigm helps streamline user profiling and classification, it also imposes oversimplifications for open-ended interfaces.
Crucially, it inherently assumes \textit{homogeneity}: two ``six-month users'' have comparable mindsets and repertoire for the system.
Thus, more experience approximates more convergence towards the ``expert'' understanding.

However, natural language interfaces take versatile inputs and allow distinct interaction modes.
This makes the experience \textit{path-dependent}~\cite{luchins1942mechanization,mittone2024initial,Wiedenbeck2004FactorsAC}: similar exposure does not entail similar understandings or skills, but each individual's decisions are rather determined by their own trajectories of learning.
Recent HCI work has begun to systematically describe the heterogeneity. For instance, \citet{BarkeCopilot} considers the different grounding at individual level when a programmer starts to adopt GitHub Copilot.
Nonetheless, most research to date continues to model users' past experience with LLMs as coarse-grained categories.

Another major thread of recent research considers Human-AI Interaction as a \textit{longitudinal} process~\cite{kjaerup2021longitudinal,thomson2003hindsight}, where both the system's and the users' behavior patterns evolve over time~\cite{long2025facilitating,qian2020whydifficult,zhu2025data,zhu2025requestmaking}.
This helps to incorporate developmental dynamics that shape long-term usage, and extends the common single-session study setup.
Yet, relevant HAI studies are subject to the known limitations of longitudinal experiments, such as limited participant numbers and crowd loyalty~\cite{thomson2003hindsight,soprano2024loyalty,abbad2016analyzing}.
Besides, existing implementations usually focus on specific downstream applications such as journalist support~\cite{wang2025role}, storytelling~\cite{Fabre2025storytelling}, or scientific communication~\cite{long2025utility}.

Our paper complements current understandings by introducing quantitative insights specifically for the trajectories of real-world LLM users, and examines the more fundamental mechanism of longitudinal patterns, e.g., whether a user chooses to stay over time, and what interaction patterns \textit{across tasks} are developed.

\subsubsection{User-LLM Interaction data From the Wild}
\label{subsubsec:RelatedWork_from_the_wild}

Recent HCI and UX research has started to consider whether and how system design apply to real-world contexts~\cite{chamberlain2012wildresearch}, grounding AI/LLM from a lab artifact into more unpredictable and complicated real user scenarios.
\citet{sun2024genaiwild} collects GenAI users' insights into the prospects and challenges in co-creating, underscoring the deeper ``uncertainties and complexities arising from resource availability'' amid the ``useful GenAI'' narrative.
\citet{shaikh2025creating} proposes to create \textit{General User Models}, which proactively take unstructured observations or past interactions across time and domains as input.
Most relevantly, \citet{zhu2025data} discusses utilizing user-LLM interaction data in the wild, pointing out that large-scale chat logs are unique as the boundary between \textit{data} and \textit{interaction} is growing increasingly blurred but currently facing usability challenges.

Understanding interaction in its natural state is not only useful but necessary for user studies and design, as key aspects of user experience are often unknown beforehand and need to be constantly updated from real-world feedback.
For instance, ~\citet{shankar2024validates} describes a ``criteria drift'': users' standards for quality and their understanding of system capabilities continuously and inevitably shift based on model outputs.
This represents the co-adaptation procedures crucial to real-world deployment that need to be re-gathered with the real run-in and adaptation of users~\cite{schneider2025mental}.
~\citet{wang_end_user_2025} discuss the lack of support for \textit{rapid initialization} and \textit{easy iteration} of contemporary social media curation tools, which fail to recognize real users' limited attention and motivation in fragmented sessions.
Beyond HAI, our work also shares high-level insights with \citet{brown2023usinglogs}, which considers understanding software engineers' work logs as a sustainable, non-intrusive text-based method for interpreting their flow and focus time.

This paper is based on a cornerstone recent work among the first to collect user chat logs as a manageable resource: WildChat~\cite{zhaowildchat,linwildbench}, a corpus of one million user-ChatGPT conversations collected via opt-in consent on Hugginface from April 2023 to May 2024\footnote{The design of our work in fact takes advantage of its featured punctuality, such that the real-world data reasonably resembles authentic early explorations with limited to no experience and knowledge.}.
In addition to full transcripts, WildChat is enriched with metadata such as geographical information, request headers, and hashed IPs, enabling analyses of user behavior across time and locale.
One contribution of WildChat is to expose a broad diversity of prompt patterns (ambiguous requests, code‐switching, etc.), making it a rich resource both for fine-tuning instruction-following models and for studying safety/risks in the wild.
Aside from WildChat as most suitable for this work, we also note other resources:
~\citet{zhenglmsys} introduced LMSYS-Chat-1M, a million‐scale dataset of real-user conversations collected in the wild via the Vicuna demo and the well-known Chatbot Arena platform~\cite{chiang2024chatbot} across 25 models and up to 210K unique IP addresses,
%However, as an earlier by-product of the Chatbot Arena, also 
but it lacks a complete user protocol and is limited in metadata~\cite{zhu2025data}.
ShareGPT~\cite{ShareGPT52K} and BIDD~\cite{trippas2024bard} are also similar useful resources but limited in scale.

\paragraph{Task-Independency} Accompanying the real-world user data is the dramatic variety in form and content.
The dominant paradigm in HAI research employs task-specific methodologies, with the focus on highly specific task-completion processes in target scenarios, such as ``students using LLMs for school''~\cite{ammari2025studentsreallyusechatgpt,10.1145/3706598.3713714,zhang2025navfog}.
Yet, with more awareness of naturalistic, real-world use~\cite{tossell2012getting}, technology should not only be viewed as a tool for discrete tasks, but also as a medium with rich information regarding users and interactions per se~\cite{dourish2001action}.
In the context of our work, we focus primarily on form: \textit{how} a user asks rather than \textit{what} they are asking. We will use an annotated variation of WildChat from Natural Language Processing, namely ReCCRE~\cite{zhu2025requestmaking}, to supply the real-world heterogeneity of use scenarios and evaluate \textit{expressions} in interactions with LLMs \textit{across different tasks}.

\section{Dataset and Experimental Setup}
\label{sec:dataset_and_setup}

\begin{figure}[t]
    \centering
    \includegraphics[width=0.42\textwidth]{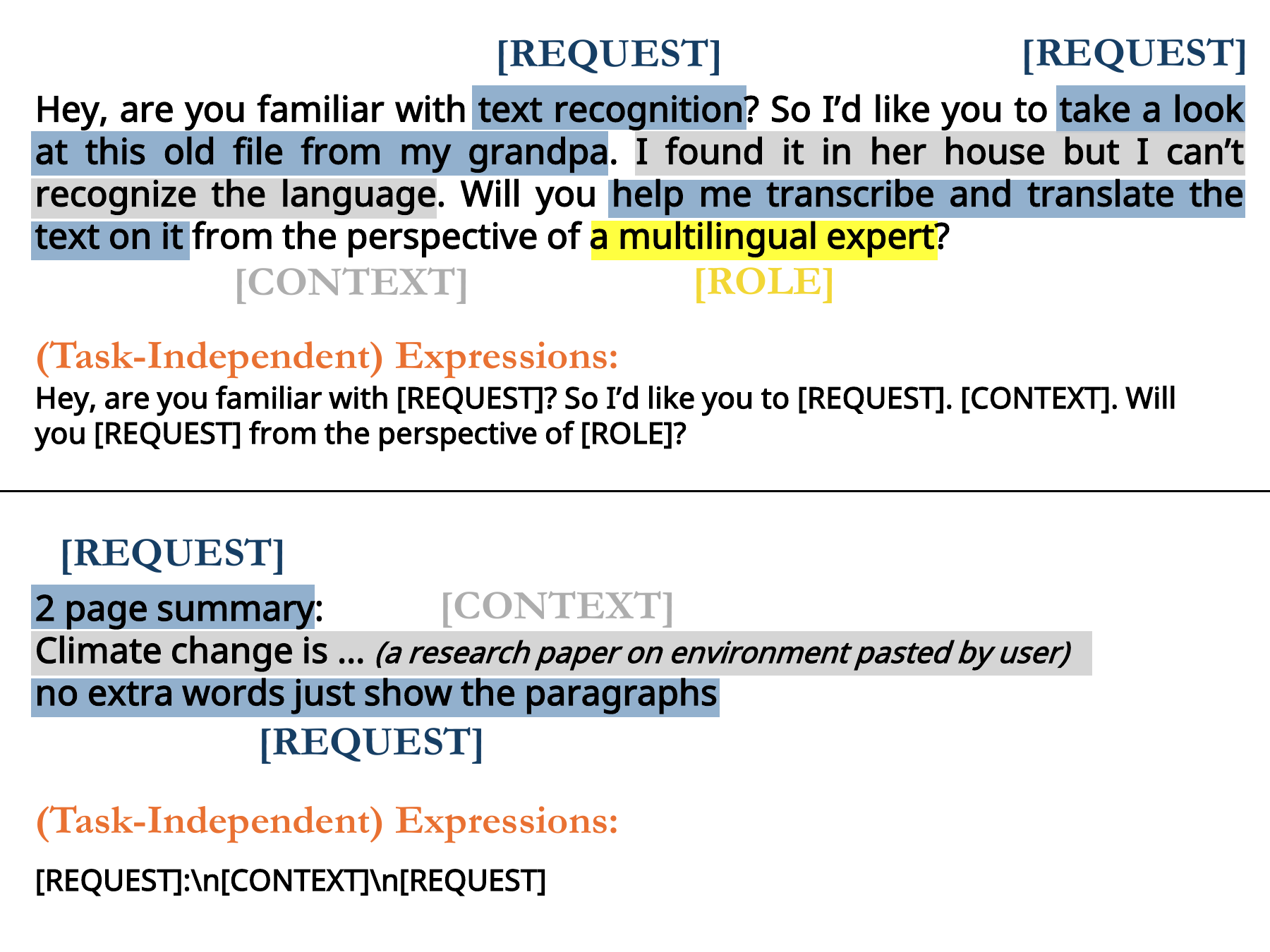}
    \caption{Two examples of ReCCRE~\cite{zhu2025requestmaking}, a dataset that performs the division of (task-independent) expressions, request/task content, and other components \textit{(content and style adapted from \citet{zhu2025requestmaking})}.}
    \label{fig:reccre_example}
    \Description{Examples of how the ReCCRE dataset segments a user input into ``request'', ``expressions'', and other components include ``context'' and ``role''. The original text reads: ``Hey, are you familiar with text recognition? So I’d like you to take a look at this old file from my grandpa. I found it in her house but I can’t recognize the language. Will you help me transcribe and translate the text on it from the perspective of a multilingual expert?'' The extracted requests include ``text recognition'', ``take a look at this old file from my grandpa'', and ``help me transcribe and translate the text on it'', and the task-independent expression remaining is ``Hey, are you familiar with [...] ? So I’d like you to [...]. [...]. Will you [...] from the perspective of [...]?''}
\end{figure}

\begin{figure*}[t!]
    \centering
    \includegraphics[width=0.85\linewidth]{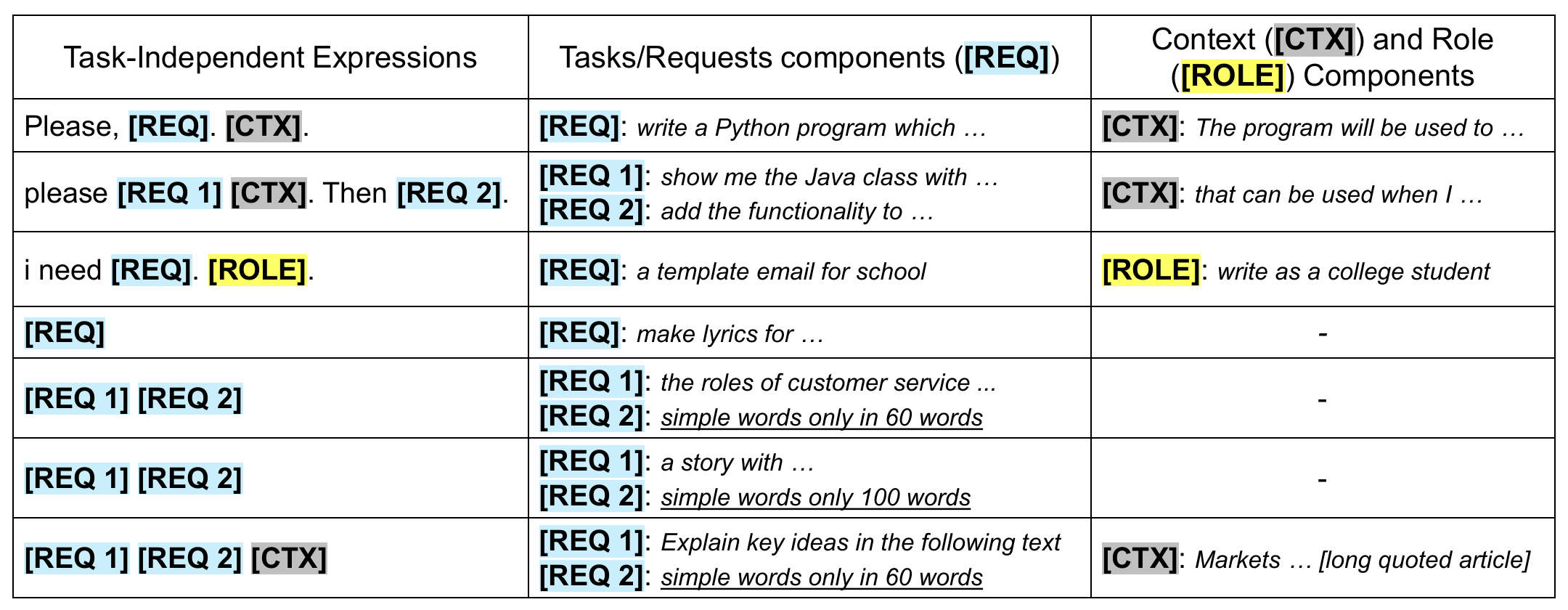}
    \caption{An example from a real-world user's chat logs. The user created 7 sessions total, spanning multiple tasks and scenarios like coding, academics, and creative writing. This figure shows how the ReCCRE dataset arranges the user inputs as standalone sections, and this paper mainly utilizes the task-independent expression patterns and the request content. We note some preliminary and representative ``molding'' procedures in the user's expressions (dropping some patterns like ``please'' and switching to direct concatenating requests) as well as in some request components (the ``simple words only'' in different tasks).}
    \label{fig:user_example}
    \Description{An example from a real-world user's conversation records. The user explored expressions including ``Please, [...].'', ``I need [...]'', and simply ``[...] [...]''.}
\end{figure*}

\subsection{Data}
\label{subsec:dataset_basics}

Our work is based on ReCCRE~\cite{zhu2025requestmaking}, a processed subset of WildChat~\cite{zhaowildchat} that selects 211,414 sessions from April 2023 to May 2024, where the user's first-turn input is proposing a ``request'' or ``task'' to be fulfilled by the LLM (as opposed to, e.g., a direct information-retrieval question like ``who is the P.M. of Canada'').
Our analyses focus on the \textit{initial prompts} of each dialog session. This is not only based on ReCCRE data availability but also a key design: the initial prompt is crafted by users with minimal constraints, reflecting their own language use and original intentions.
By aligning this space of initial inputs, we can interpret how users initiate dialogs and deliver requests, before the full dialog becomes too heterogeneous and context-dependent.
This initial input is interchangeably referred to as the \textit{first turn}, following terminology of conversation studies.

Each dialog session in ReCCRE/WildChat is documented with its metadata, including key attributes relevant to our analysis:
\begin{itemize}
    \item{\textbf{Anonymized IP}}. We use this information to group the dialogs from the same user.
    \item{\textbf{Timestamp}}. A user's dialogs are arranged chronologically to study interaction over time.
    \item{\textbf{Model}}. WildChat's interface provides two entrances with the same UIs for users, for chatting with a GPT-3.5 and a GPT-4 (available after October 2023) model respectively. We observe users' transition from GPT-3.5 to GPT-4 after the latter is up, as a natural ``intervention'' (\S\ref{subsubsec:Experiments_model_change_yes_expression_diff}).
\end{itemize}

We follow the notion of \textbf{long-term users} or \textbf{core users} defined in ReCCRE, referring to users with activity spanning at least 15 days and having created at least 10 dialog sessions. The prolonged and steady involvement with the system enables meaningful conclusions regarding procedural and longitudinal patterns. In practice, the core user group consists of 2,032 global users with 59,017 sessions created.
All remaining users are referred to as ``non-core'' or ``other users''.
As our work focuses on the change across dialog sessions over time, we further filter out users with too few activity records to form a meaningful exploration process ($\le3$ sessions total).
This narrows down to 5,923 non-core users and 80,518 sessions. The combination of the two groups is referred to as ``all users'' (7,955 users, 139,535 sessions). The median number of sessions is 14 for core users and 8 for all users.
To ensure the most rigorous setup for conclusions about longitudinal interactions, and following the recommendations of ReCCRE, experiments are by default performed with the core users. However, for any part that concerns the contrast between user groups or the length of user lifecycles (\S\ref{subsubsec:Experiments_no_convergence_across_users}, \S\ref{subsec:Experiments_early_interaction}, \S\ref{subsubsec:Experiments_model_change_yes_expression_diff}), we cover all users for a more universal understanding.

\subsubsection{Divison of Components}
Real-world LLM data varies substantially in form and content. 
The distinguishing feature of ReCCRE is the division of different \textit{components} in any request-making utterances, as exemplified in Figure~\ref{fig:reccre_example}. Specifically, it implemented a generalizable taxonomy for any request-making utterances in natural or human-LLM conversations with the following elements (\textit{italic: quoted directly from \cite{zhu2025requestmaking}}; normal text: additional notes on how this work uses the component or not):
\begin{itemize}
    \item{\textbf{Request Content}}: \textit{``A core span of text that specifies what task(s) exactly the user wants the LLM to perform, or what goal(s) the user wants to achieve.''}
    \item{\textbf{Context}}: \textit{``A detailed span of context information that does not directly constitute the request, but provides support for neighboring requests.''} Our work does not consider the context components due to their overwhelming heterogeneity and uncertain relation with the requests (anything that users may quote, paste, or refer to, such as the article to be summarized after the command of ``Summarize this:'').
    \item{\textbf{Role}}: \textit{``Any roles that the LLM is asked to take on to achieve the requests.''} Roles are not a major focus due to their practical scarcity: while the request contents and expressions exist in almost any data points, less than 10\% of the sessions in the dataset contain role assignment. % \footnote{Yet, we show in Fig.~\ref{fig:conv_role_cosine} that the Role component shares a similar trend of converging.}
    \item{\textbf{Task-Independent Expressions}}: \textit{``The remaining text after extracting the above components, which represent the generic language templates used to embed and deliver the requests.''} This is the core of our study as it represents the exact way that users interact with LLMs.
\end{itemize}
While the dataset was originally proposed for pragmatics study of request-making behaviors, we use it as the base of our research, as separating task-independent expressions from task-specific content is central to a fundamental understanding of real-world user-LLM interaction.
By isolating the reusable expression ``templates'' users employ, we can pivot the focus to how users have learned to initiate interactions.
Identifying these general expression patterns allows us to observe the formation of underlying patterns directly, and they provide a powerful signal for their mental model, interactional habits, their levels of fluency with the system, or when a user has settled into certain paths.

\subsubsection{Overview of the research setup from a real-world example of user expression formation}
Figure~\ref{fig:user_example} depicts the documented activities from a real-world user's full records of 7 sessions in chronological order.
This example provides a preview of this work's goals and representative patterns of user input.
The user's requests cover a broad range of tasks, including coding, creating writing (lyrics, stories), communications (email), and academic/professional usages.
With the ReCCRE format, we can make sense of the interactions from expression templates across content.
Here, the user tends to input brief task instructions stacked in minimalist ways; upon their last few attempts, they interact with the LLM by directly giving request components without wrapping them in more conversational formatting or other expression patterns.
This procedure captures the user's explorations and changes of expressions, including the dropping of early patterns (politeness marker ``please'', auxiliary conversation setups ``I need ...'', and punctuations) and testing out a role assignment.
Besides, we also notice an auxiliary [REQ], \textit{``simple words, only [num] words''}, that becomes a common request component to include as the user switches to the more ``direct'' expressions.
This paper seeks to capture these changes over time, and provide a set of quantitative analysis generalizable across users to provide more insights into such procedures. We note especially the more fundamental task-independent expressions that create the very form of real-world user-LLM interaction but are less studied.

\subsection{Experimental Setup}
\label{subsec:experimental_setup}

\subsubsection{Representing Texts as Vectors}
To computationally model the components of user inputs, we employ text embedding to project them into a high-dimensional vector space. This transforms variable-length text into fixed-length, dense vector representations where the geometry of the space encodes semantic relationships. Texts with similar meanings are thus positioned proximally within this space. Specifically, we utilize \texttt{gte-large-en-v1.5}, a model from the General Text Embeddings (gte) family~\cite{zhang2024mgte,li2023towards}. This model series was specifically selected due to its state-of-the-art performance on the Massive Text Embedding Benchmark (MTEB)~\cite{muennighoff2023mteb} for semantic encoding tasks, ensuring a robust and nuanced capture of both the structure and meaning of user inputs.

\subsubsection{(Dis)similarity as metric of distances between sessions}
The majority of our experiments are based on \textit{text (dis)similarity} between two user inputs, measured with \textit{cosine similarity} calculated between two vectors, which compares the direction of vectors after normalizing to length 1.0. In practice, we use the inverse of cosine similarity, i.e., the \textit{cosine distance} metric $d(c_1, c_2) = 1 - sim(c_1, c_2)$. A larger number indicates a greater difference between the pair of texts, and vice versa. A minimum value $0$ is reached when the two input texts are identical. This allows us to robustly quantify the dissimilarity across user inputs and expressions in our dataset.

\section{Experiments \& Findings}

We first set up the definitions of the ``molding'' process in question and validate its common presence. We note the different dynamics of the request and expression components and the nuances. Later, we show that the early interactions are highly associated with a user's longitudinal behaviors including textual patterns and long-term retention. Finally, we suggest two key factors, task stratification and model updates, as representative examples of compelling dynamics, and examine how they add to the general molding.

\begin{figure*}[t]
    \centering
    \subcaptionbox{k=1}{%
        \includegraphics[width=0.28\textwidth]{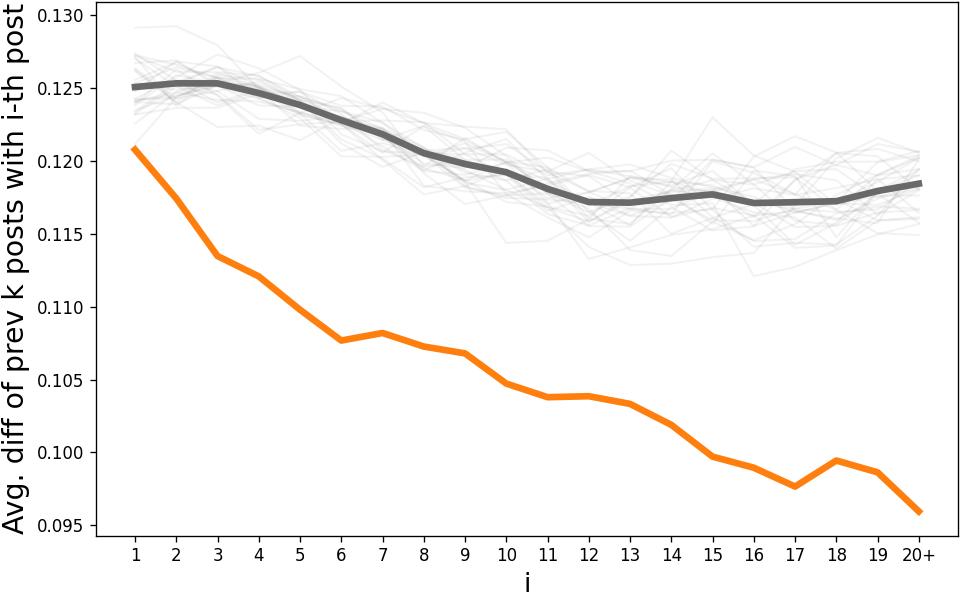}
    }
    \subcaptionbox{k=3}{%
        \includegraphics[width=0.28\textwidth]{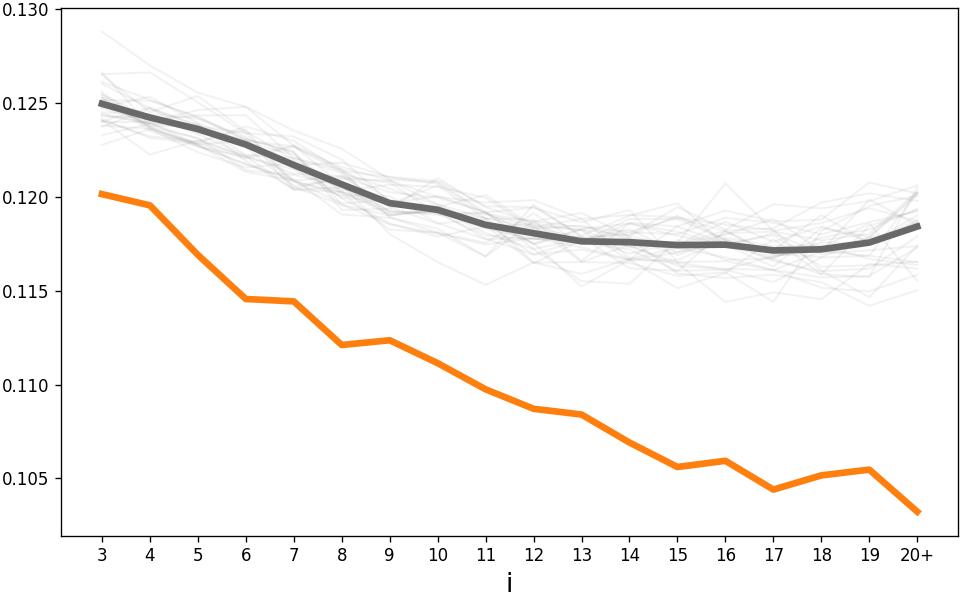}
    }
    \subcaptionbox{k=5}{%
        \includegraphics[width=0.28\textwidth]{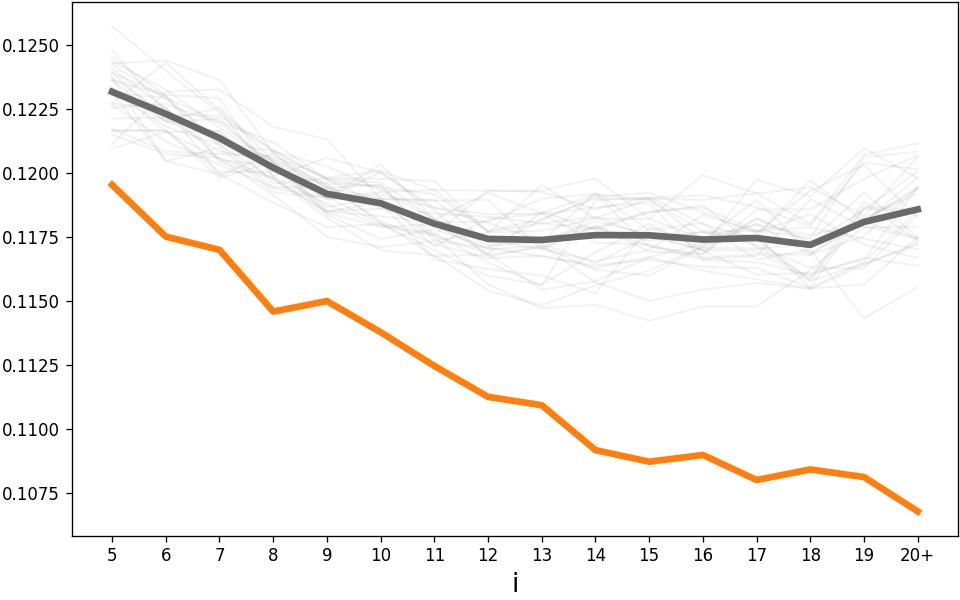}
    }
    \caption{Convergence by Recency: The difference between adjacent ($k=1,3,5$) \textbf{expressions} becomes much smaller with more sessions (orange line), compared to the same sessions randomly shuffled (bold grey line). Each of 50 thin gray lines represents a random trial shuffling all sessions for each user, and aggregated as the bold gray line. Similarly for Fig.~\ref{fig:conv_request_cosine} and \ref{fig:conv_role_cosine} below.}
    \label{fig:conv_expression_cosine}
    \Description{The average cosine distances for the expression component with window size 1, 3, and 5. As the sessions increase, this distance of adjacent sessions decreases smoothly and gradually from around 0.12 to 0.095, while the random shuffle baseline stays around 0.12.}
\end{figure*}

\begin{figure*}[t]
    \centering
    \subcaptionbox{k=1}{%
        \includegraphics[width=0.28\textwidth]{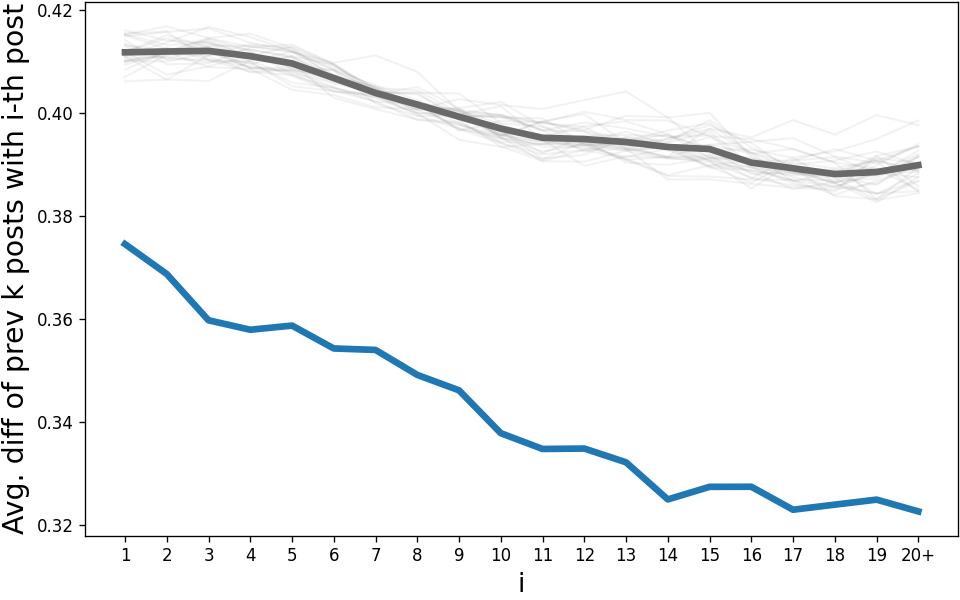}
    }
    \subcaptionbox{k=3}{%
        \includegraphics[width=0.28\textwidth]{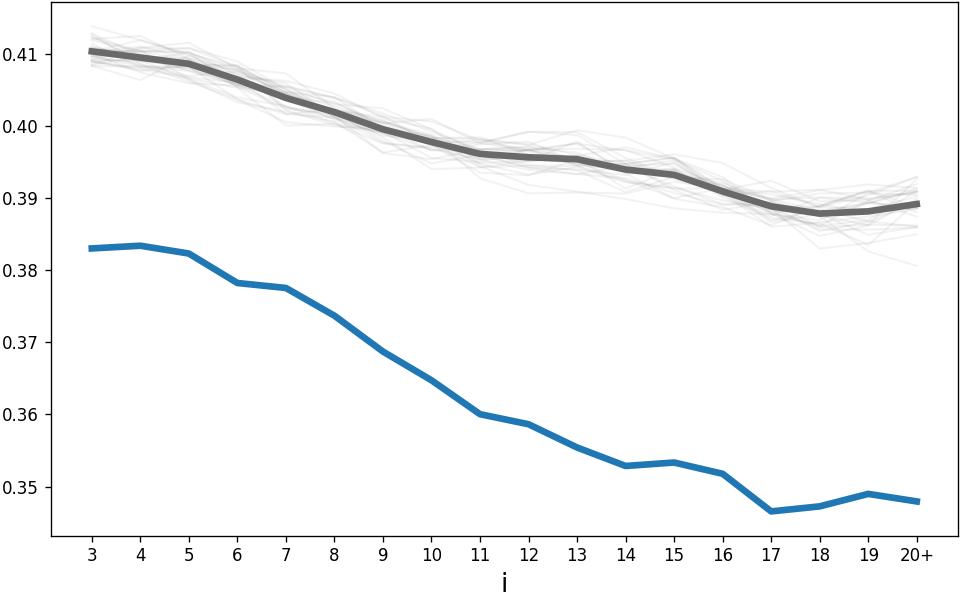}
    }
    \subcaptionbox{k=5}{%
        \includegraphics[width=0.28\textwidth]{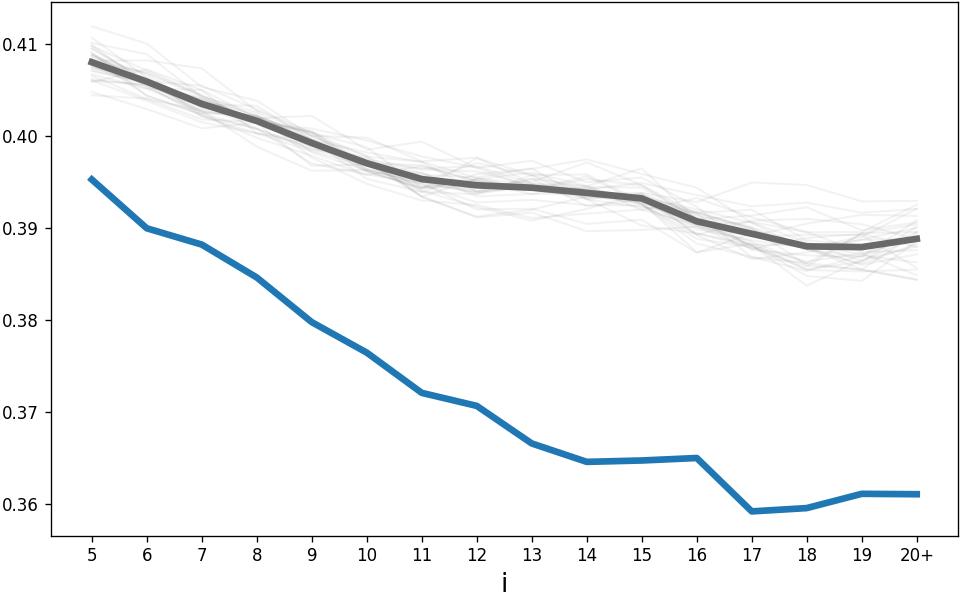}
    }
    \caption{The decreasing difference between adjacent ($k=1,3,5$) \textbf{requests} with more sessions (blue line).}
    \label{fig:conv_request_cosine}
    \Description{The average cosine distances for the request component with window size 1, 3, and 5. As the sessions increase, this distance of adjacent sessions decrease smoothly and gradually from around 0.38 to 0.32, while the random shuffle baseline stays around 0.4.}
\end{figure*}

\begin{figure*}[t]
    \centering
    \subcaptionbox{k=1}{%
        \includegraphics[width=0.28\textwidth]{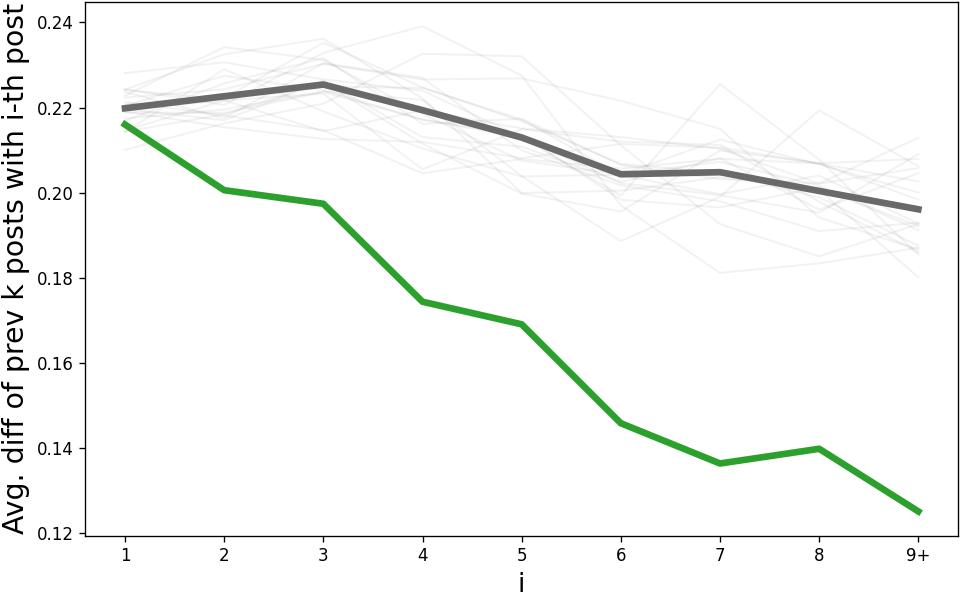}
    }
    \subcaptionbox{k=3}{%
        \includegraphics[width=0.28\textwidth]{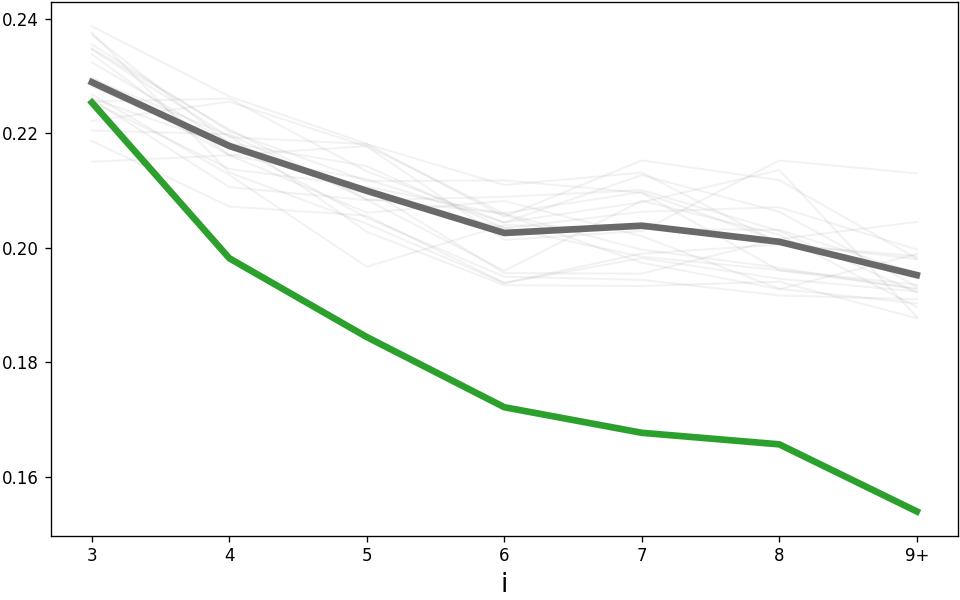}
    }
    \subcaptionbox{k=5}{%
        \includegraphics[width=0.28\textwidth]{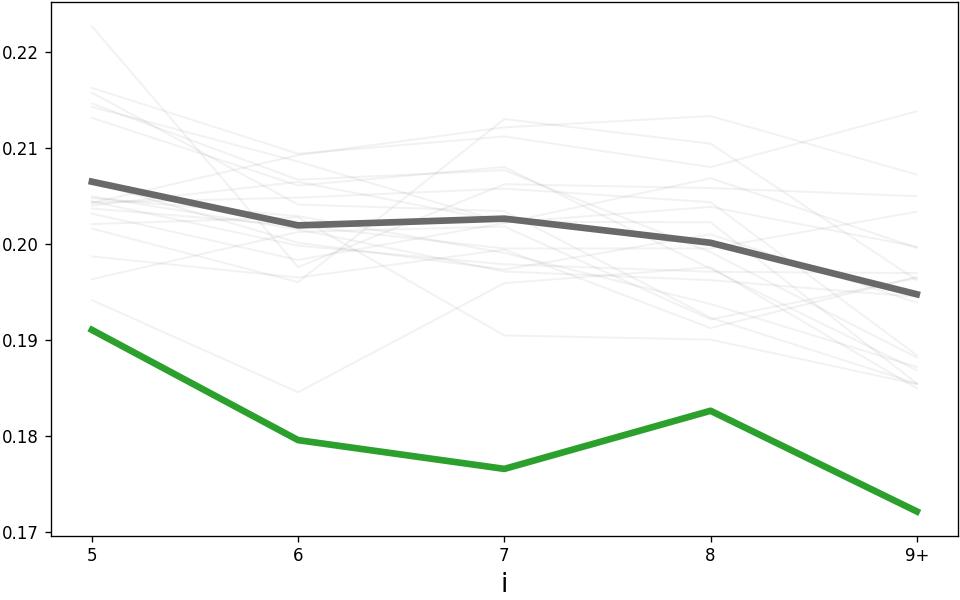}
    }
    \caption{The decreasing difference between adjacent ($k=1,3,5$) \textbf{roles} with more sessions (green line).}
    \label{fig:conv_role_cosine}
    \Description{The average cosine distances for the role component with window size 1, 3, and 5. As the sessions increase, this distance of adjacent sessions decrease smoothly and gradually from around 0.22 to 0.12, while the random shuffle baseline stays around 0.22.}
\end{figure*}

\subsection{The molding of user input}
\label{subsec:molding_user_input}
Initial work on the ReCCRE dataset~\cite{zhu2025requestmaking} finds that, on average, a user's expressions in their latest inputs appear less and less different from the previous sessions. 
We start by systematically validating and extending this direction, taking into consideration all components and variables of user input.

Consider the initial inputs of all dialog sessions from a user $u$ in chronological order, denoted as $C^u = \{c^u_1, c^u_2, ..., c^u_n\}$. %, or abbreviated as $C =\{c_i, \forall i\}$ when unambiguously referring to the one user in question.
We use two complementary metrics to quantify the convergence of a user's inputs based on the distances between the user input in the $i$-th dialog and the preceding dialogs:
\begin{itemize}
\item{\textbf{Variation}} compares a new input to all past sessions. We define it as the distance to $c_i$'s closest prior: $\min d(c_{i}, c_{j})$, $\forall j\in\{1, 2, ..., i-1\}$.
This is a \textit{global} measurement of dispersion in a user's input space, reflecting the total degree of their exploration.
We expect this value to drop as the total number of sessions increases.
\item{\textbf{Recency}} compares a new input to the $k$ most recent sessions. This measures \textit{locally} whether a user is converging to similar patterns, as their expressions or requests see fewer changes. Formally, we measure Recency as $\overline{d(c_{i}, c_{i-j})}$, $\forall j\in\{1, ..., k\}$, i.e., the average distance with a given window size $k$.\footnote{The original ReCCRE~\cite{zhu2025requestmaking} considers the \textit{minimal} distance of a new input with the prior $k$ sessions. This is essentially a slightly weaker version of Recency. For instance, suppose $c$ is periodical with loop length $k$, s.t. $c_i \equiv c_{i-k}$. Thus $\min d(c_{i}, c_{i-j}) \equiv 0$ when $j=k$. However, the Recency here as defined by \textit{average length} can be arbitrarily large given $c_{i-k}$ to $c_{i-1}$.}
    
\end{itemize}
Despite the close relations between the two, we will show that convergence in the recent window does not entail a bounded variation, which differentiates users' request content and their expressions.

\begin{figure*}[t]
    \centering
    \subcaptionbox{request (linear, $R^2 = 0.996$)\label{subfig:total_type_num_request}}{%
        \includegraphics[width=0.28\textwidth]{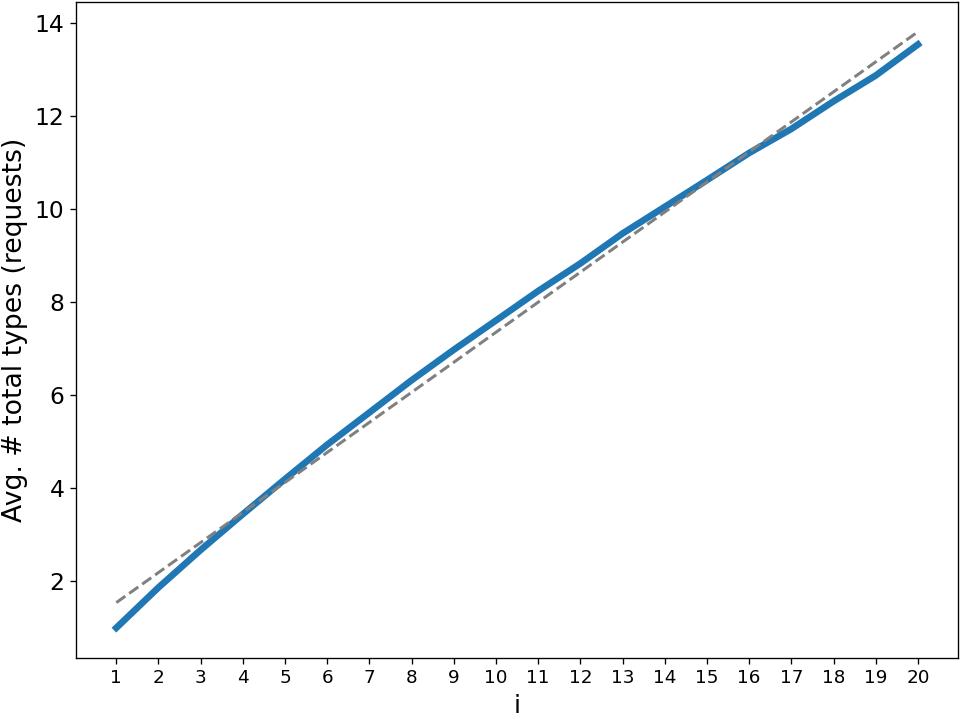}
    }
    \hspace{2pt}
    \subcaptionbox{expression (logarithm, $R^2 = 0.997$)\label{subfig:total_type_num_expression}}{%
        \includegraphics[width=0.28\textwidth]{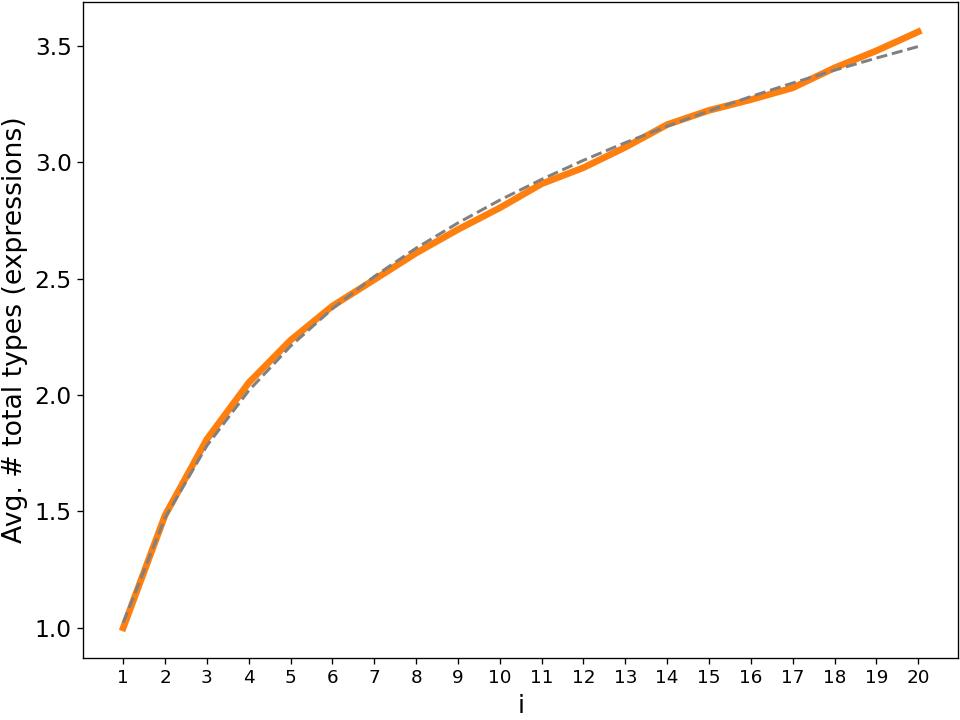}
    }
    \hspace{2pt}
    \subcaptionbox{scale comparison\label{subfig:total_type_num_scale-comparison}}{%
        \includegraphics[width=0.28\textwidth]{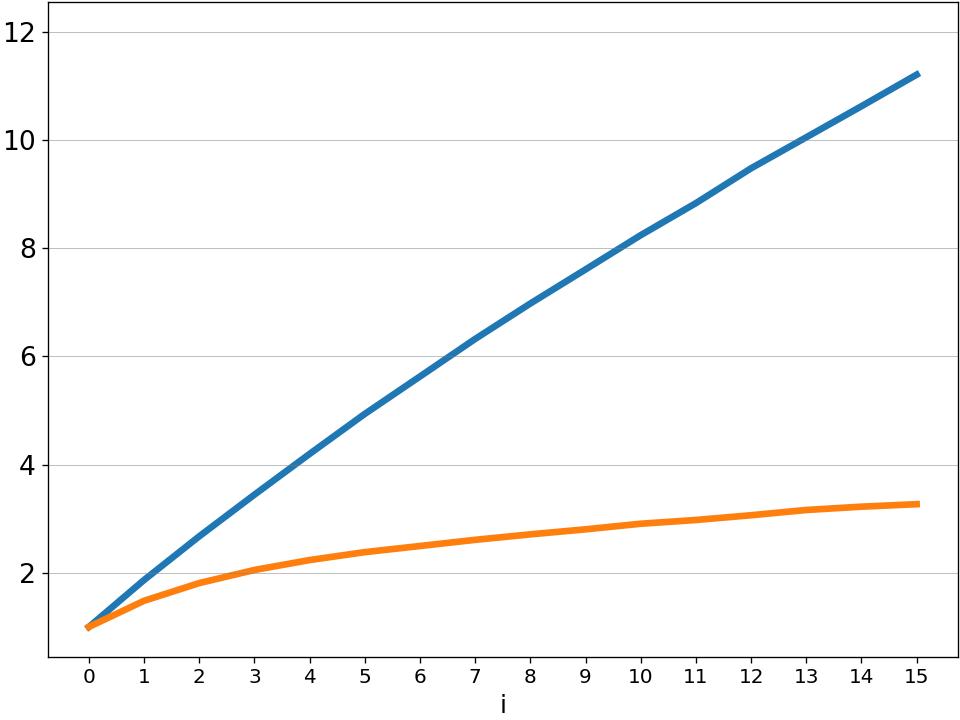}
    }
    \caption{The distinct dynamics behind the molding of expressions versus requests. Users are constantly trying new requests: the number of request types increases linearly with the session count (\ref{subfig:total_type_num_request}). In contrast, the number of expressions grows logarithmically: users rarely try new expression types after their initial experience (\ref{subfig:total_type_num_expression}). The scales are compared in \ref{subfig:total_type_num_scale-comparison}; in fact, users stick to a handful (2 to 4 different types) of expressions after just a few sessions.}
    \label{fig:total_type_num_and_fit}
    \Description{The constant linear increase of total type numbers for requests, and the logarithmic increase for expressions.}
\end{figure*}

\begin{figure}[t]
    \centering
    \includegraphics[width=0.4\textwidth]{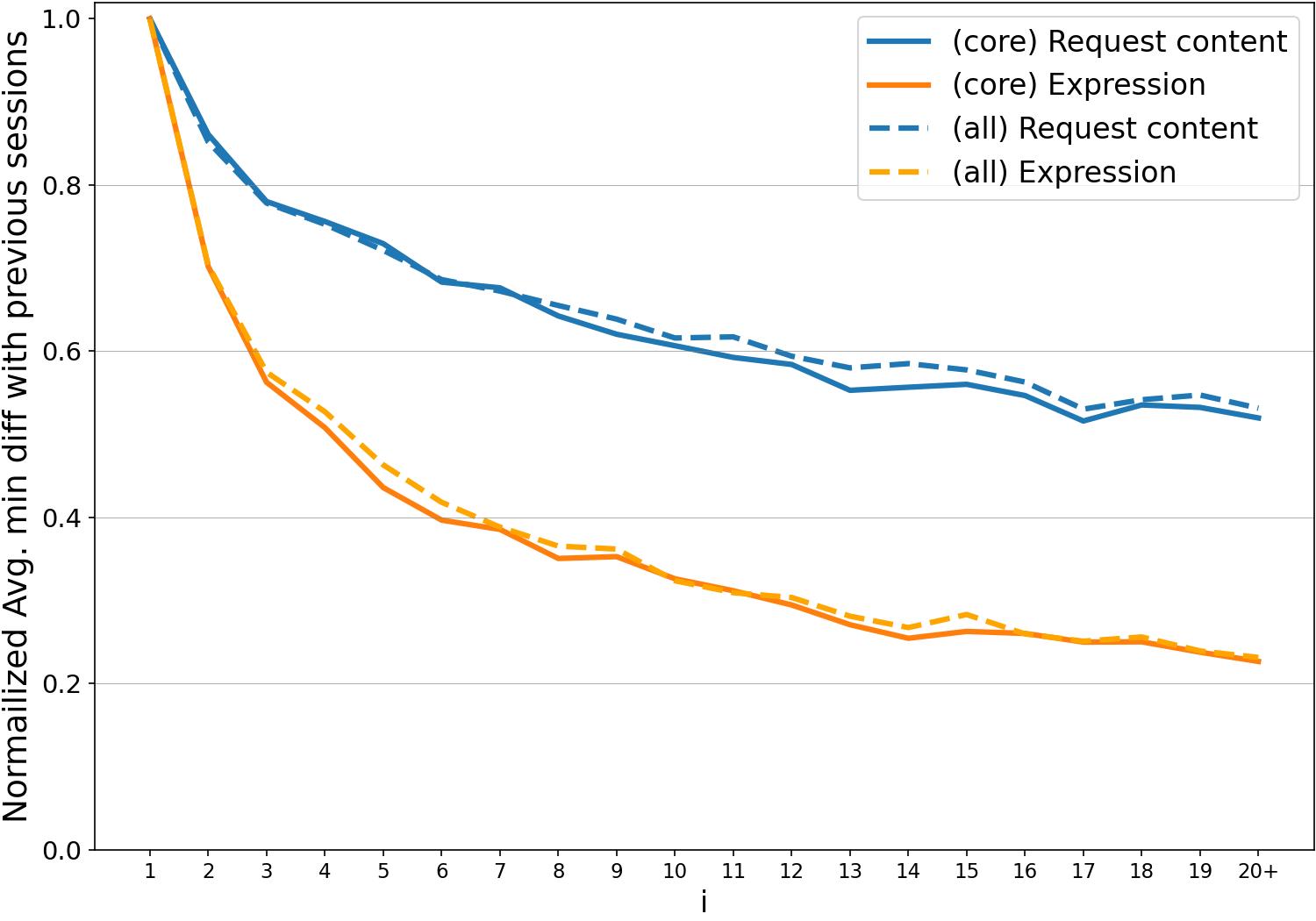}
    \caption{Variation, the (normalized) distance to the single most similar previous turn, drops for both request components (blue lines) and expression components (orange lines); however, expressions become similar to existing sessions significantly more quickly, and the decrease in distance is to a much more notable degree ($\sim$$20\%$ remaining after 20 sessions). This trend is consistent for both core users (solid lines) and all users (dashed lines).}
    \label{fig:variation_normalized}
    \Description{The (normalized) distance to the single most similar request and expression components. Request Variation dropped to 60\% in around 10 sessions, but then stabilized around 50-60\%. Expressions, however, dropped to 60\% in only 3 sessions, to 30\% in 10 sessions, and keeps decreasing to around 20\% in 20 sessions.}
\end{figure}

\subsubsection{All components of a user's input converge over time}
We first confirm that not only task-independent expressions but in fact all components, including request content and roles assigned, are molded as users interact with the system more and more.
Figures~\ref{fig:conv_expression_cosine}, \ref{fig:conv_request_cosine}, and \ref{fig:conv_role_cosine} respectively show that distances decrease (molding) under cosine dissimilarity as users interact with the system, for all three components: expression, request/task content, and role assignment.
At the \textit{i}-th turn, represented by the X-axis, we calculate the convergence under Recency ($\overline{d(c_{i}, c_{i-j})}, \forall j\in\{1, ..., k\}$), and obtain the average of all applicable users ($\ge i$ total sessions) as the value on the Y-axis.
Each configuration is tested under 3 different choices of window sizes $k$ to ensure robustness.
We observe a continuous decline in all components across all configurations, showing the results are not dependent on the choice of metric or parameters.

To show that the effect is indeed a result of sequential user exploration rather than simply an artifact of adding more sessions, we follow \citet{zhu2025requestmaking} and compare the convergence with randomization: The bold grey line shows the average of 30 trials (each denoted by a thinner grey line) with the dialog sessions randomly shuffled and all other setups unchanged. We see hardly any comparable tendency of convergence in this control scenario.

\subsubsection{Tasks increase linearly, expressions increase logarithmically}
\label{subsubsec:bounded_expression_types}
The previous section demonstrates that all components in a user's input become increasingly similar as time progresses. Next, we examine if there are qualitative differences between the convergence of different components. 
%Earlier we defined the notion of Variation ($v_{i} = \min d(c_{i}, c_{j}), \forall j\in\{1, 2, ..., i-1\}$) as a global measurement of convergence.
Intuitively, as a user's inputs get molded, the Variation $v_i$ will grow sufficiently small with fewer or no new inventions of expressions or tasks.
The question is how fast this is achieved --- Do task content and task-agnostic expressions share the same global model?
We discern their distinct dynamics over time, suggesting that users apply a limited set of language as interaction schemes for varied request types.

Figure~\ref{fig:variation_normalized} compares $v_i$ for \textit{task content} and \textit{expression} over time, normalized to the same scale of 0 to 1 for comparison with their initial status.
Both components experience a notable drop as users create more sessions; however, user expressions grow similar to past sessions at a much higher speed and to a much higher extent, with the distance shrinking to only 20 to 30$\%$ after 15--20 sessions. The trends for both elements are consistent for both core users (shown in solid lines) and all users (dashed lines).
This difference suggests that input tasks are continuously renovated despite the ``molding'' in the local view, but user expressions seem to tell a different story.
%Next, we seek finer-grain tools to dig deeper into this contrast, and build formal models for the different dynamics.

\paragraph{A handful of \textbf{Task-Independent} user expressions}
%Consider the total number of \textit{different types} of a component.
For each user, we consider the total number of \textit{different types} of a component.
For each new input $c_i$, we categorize it as either a similar instance of a previous type (if its variation $v_i$ is less than a threshold $\delta$ from an element of the set of types) or add it to the set as a new type.~\footnote{We show in Appendix~\ref{appendix:params} that the findings here are insensitive to the choice of $\delta$ (within reasonable range), consistent with no qualitative difference as $\delta$ ranges from small (0.075) to huge (0.175). Figure~\ref{fig:total_type_num_and_fit} uses $\delta=0.125$.}
%The variable $j_i$ indicates the index of the type input for the $i$th input, such that if $j_i=i$ the $i$th input is the first instance of a new type.
%, we view it as the same type as its \textit{most similar prior} $j_i = \arg\min_j [d(c_{i}, c_{j})]$. Otherwise, we call $v_i$ a new type with the most similar prior defined as itself ($j_i = i$). Thus, $j_i$ is a direct indicator of how a user's input evolves: the ``crystallization'' of a component would expect a slow increase or even convergence of its corresponding $N_{type}$.

\begin{figure*}[t!]
    \centering
    \subcaptionbox{Avg. distance \textit{within} a user\label{subfig:dist_within_user}}{%
        \includegraphics[width=0.36\textwidth]{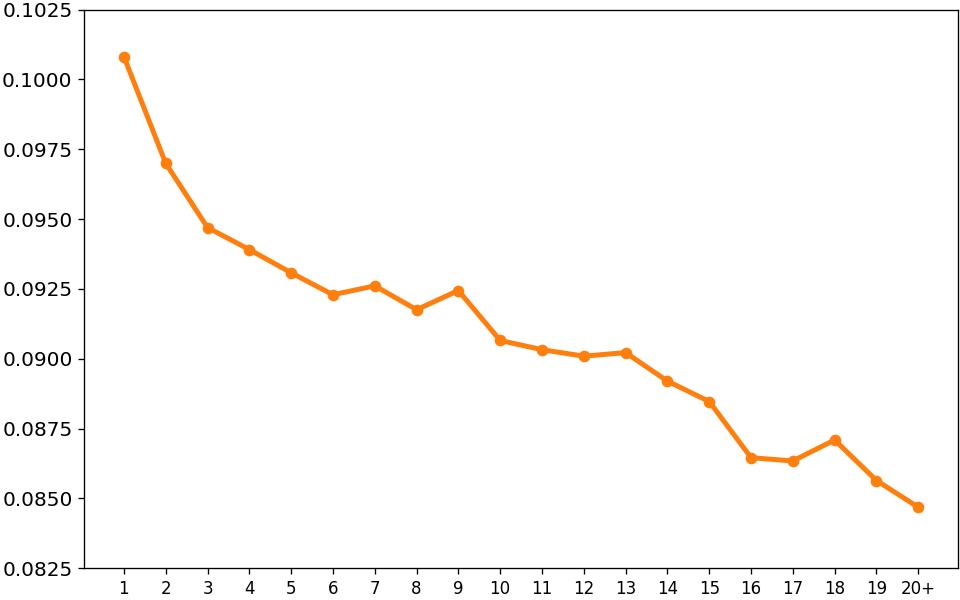}
    }
    \hspace{9pt}
    \subcaptionbox{Avg. distance \textit{across} users\label{subfig:dist_across_users}}{%
        \includegraphics[width=0.36\textwidth]{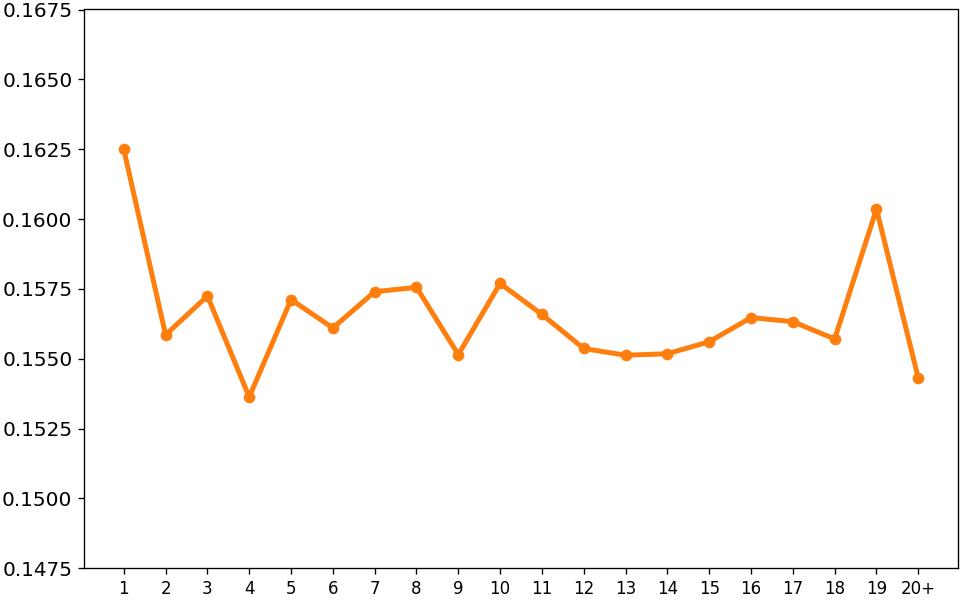}
    }
    \caption{Comparison of the average distances for expressions within the same user's trajectory (\ref{subfig:dist_within_user}) and between different users (\ref{subfig:dist_across_users}) at the $i$-th sessions of all users, displayed under the same scale. With more sessions, the within-user distances continue to decrease, but this trend is not present across users, indicating that the users are not converging to the same expressions.}
    \label{fig:dists_across_users_no_convergence}
    \Description{The distances within the same user gradually and steadily decreases from 0.101 to 0.094, but across-user distance fluctuates around 0.157 and hardly decreases.}
\end{figure*}

We first consider the total number of different types $N_{type}$, i.e. the cardinality of the set of types. The results (average of all core users) for the requests/tasks and the expressions are respectively illustrated as Fig.~\ref{subfig:total_type_num_request} and Fig.~\ref{subfig:total_type_num_expression}.
Interestingly, we observe an almost perfectly \textbf{linear} increase of $N_{type}$ of \textbf{requests/tasks} ($N_{type} \approx 0.645\cdot i + 0.898$, $R^2 = 0.996$).
In contrast, the total number of \textbf{expressions} quickly gets stagnant after a short period of initial exploration. In fact, it fits well to a logarithmic increase model ($N_{type} \approx 1.004\cdot \log(i + 0.768) + 0.450$, $R^2 = 0.997$).
Note especially the different scale, as compared in Fig.~\ref{subfig:total_type_num_scale-comparison}. The average number of different expressions of a user actually converges to a significantly small number (between 2--4) within only $\sim$$5$ sessions, but users continue to bring in previously unseen tasks for the LLM to process.

Paradoxically, although an open-ended LLM chat service offers an unconstrained input space for users to freely explore distinct prompts by design, users in reality show notable reluctance against variation in expressions.
Within a handful of sessions, users develop a set of as few as 2--3 ways on average to wrap up and express their requests.
This contrast reveals a gap between real-world user-LLM interaction and typical task-oriented study setups: Instead of constantly updating and adapting, users may more often appear with an ``unmotivated/low-effort'' situation~\cite{tetard2009lazy,zipf2016human}, where a finite and stable set of interaction modes accommodates various scenarios. %This is further discussed in \S\ref{subsubsec:Experiments_task_diff_no_expression_diff}, where we confirmed that different tasks are \textit{not} associated with different expressions.
% Hence, studies of behavioral patterns in distinct subfields might be subject to the same underlying bias, and it is possible that findings thereof are in fact partly describing this basal level of users.
%Meanwhile, it is also an opportunity towards a more fundamental and generic understanding of users and their behavioral patterns, which might be valuable for personalized applications and general user models~\cite{shaikh2025creating}.

\subsubsection{No convergence \textbf{across} users}
\label{subsubsec:Experiments_no_convergence_across_users}
So far we have considered each user's individual trajectory.
A natural follow-up question is the convergence \textit{across} users --- are different users molding toward the same interaction (expression) patterns after sufficient exploration?
%While the actual tasks vary dramatically, it is meaningful to explore possible consensus at expression level.
If there were universal convergence towards a final or ``optimal'' form of input language despite varied task content, it could directly guide future design with these strong natural priors.
We extend the distance measurements to consider the difference between all users' $i$-th session $\overline{d(c^{u_p}_{i}, c^{u_q}_{i})}$, $\forall u_p, u_q (p\neq q)$, checking if it also converges and comparing it against the scale and convergence speed of Recency $\overline{d(c_{i}, c_{i-1})}$.

The answer is no: Compared with the steady decrease in a user's adjacent inputs, cross-user distances do not see a similar decrease with more sessions. Further, the level of difference ($\sim$$0.156$) is significantly larger than anywhere in a user's own trajectories ($0.08$--$0.11$). Users' explorations in fact quickly amplified and solidified the heterogeneity of interaction, and they thus arrive at and keep to distinct ``local optima'' in the input space. This is in line with preliminary findings of ReCCRE~\cite{zhu2025requestmaking}: from a 2-D dimension-reduction representing each user with the average embedded expressions of all sessions a user created, it was illustrated that users seem to scatter all over the expression space with no strong convergence toward certain pattern types.

\subsection{Influence of early interaction}
\label{subsec:Experiments_early_interaction}
%We have observed the rapid formation of a user's input and the early convergence of expressions.
If the first few sessions account for the majority of a user's exploration, what do we know about this stage? Importantly, how do the early sessions associate with user behaviors in the long run? In this section, we propose two case studies.
We first consider \textbf{recurring text patterns}: common language patterns in user expressions from the first few sessions are 5 to $50\times$ more likely to be used in the remaining sessions of a user's full life cycle. 
Second, we consider \textbf{long-term retention}: prolonged activity is significantly associated with a user's early sessions; in fact, the diversity of a user's early exploration directly predicts how much longer they will stay.

\begin{table*}[t!]
\caption{Frequency of four types of keywords in the rest of a user's life cycle with/without being used in their first 5 dialogs.}
\renewcommand{\arraystretch}{1.02}
%\begin{adjustbox}{width=0.7\linewidth,center}
\begin{tabular}{V{2} c V{2} c V{2} c V{2} c V{2} c V{2} c V{2}}
\thickhline
Type & Keyword & Num ($w$/$\neg w$) & $p^{rest}_w | w$ & $p^{rest}_w | \neg w$ & Ratio \\ \thickhline
\multirow{3}{*}{Politeness} & please & 318 / 1593 & 0.29 & 0.025 & 11.5$\times$ \\ \cline{2-6} 
 & thank & 25 / 1886 & 0.041 & 0.0016 & 25.3$\times$ \\ \cline{2-6} 
 & hi,\textit{ or} hi\verb*| | & 63 / 1848 & 0.21 & 0.0042 & 49.4$\times$ \\ \thickhline
\multirow{3}{*}{Modal Verbs} & can you & 275 / 1636 & 0.23 & 0.020 & 11.5$\times$ \\ \cline{2-6} 
 & should & 54 / 1857 & 0.14 & 0.0058 & 23.3$\times$ \\ \cline{2-6} 
 & may\textit{ or} might & 27 / 1884 & 0.011 & 0.0012 & 9.3$\times$ \\ \thickhline
\multirow{3}{*}{Speech Act \& Instructions} & want & 168 / 1743 & 0.16 & 0.016 & 9.8$\times$ \\ \cline{2-6} 
 & help & 122 / 1789 & 0.17 & 0.010 & 16.3$\times$ \\ \cline{2-6} 
 & act\verb*| | & 49 / 1862 & 0.17 & 0.0041 & 39.8$\times$ \\ \thickhline
\multirow{4}{*}{Pronouns} & you\verb*| | & 563 / 1348 & 0.26 & 0.043 & 6.1$\times$ \\ \cline{2-6} 
 & i\verb*| | & 414 / 1497 & 0.23 & 0.040 & 5.7$\times$ \\ \cline{2-6} 
 & your & 31 / 1880 & 0.18 & 0.0036 & 49.3$\times$ \\ \cline{2-6} 
 & this & 68 / 1843 & 0.061 & 0.0056 & 11.0$\times$ \\ \thickhline
\multirow{2}{*}{(Baseline)} & random letter (``b'') & 379 / 1532 & 0.097 & 0.043 & 2.2$\times$ \\ \cline{2-6} 
 & ,\verb*| | & 833 / 1078 & 0.23 & 0.095 & 2.4$\times$ \\ \thickhline
\end{tabular}
%\end{adjustbox}
\Description{All keywords are significantly more often if used in early interactions, and much more significant than the trend of baselines.}
\label{tab:early_occurrence_freq_ratio}
\end{table*}

\subsubsection{Expression patterns attempted early on are significantly more likely to be reused}
%We have seen in Fig.~\ref{fig:similar_loc_distribution} that the expression component appears notably ``backtracking'' to non-recent sessions.
First, we take a closer look at the specific textual and pragmatic patterns~\cite{zhu2025requestmaking,schneider2025mental} employed by user expressions in LLM prompts.
We observe whether typical words and phrases as representative linguistic patterns appear in the first $n=5$ sessions created by the user.
This window represents a limited but decisive period in a user's earliest explorations.
Specifically, we consider four common and representative expression patterns:
\begin{itemize}
    \item \textbf{Politeness strategies and markers} directly indicate key user perceptions like social distance~\cite{brown1987politeness} with LLMs and the intended interpersonal frame of the interaction. Common markers and hedging include softening of a directive, showing appreciation, and greetings. They are strong cues for conversational studies~\cite{danescu-niculescu-mizil-etal-2013-computational} including in HAI~\cite{zhao-hawkins-2025-comparing,ribino2023role}, e.g., for styles and interlocutor relations.
    \item \textbf{Modal verbs} are important pragmatic cues~\cite{palmer2001mood,papafragou2021modality}. Encoded modality represents illocutionary force~\cite{austin1975things,alston2000illocutionary}, such as a request for action (``can you...''), normative suggestion (``should''), or signal of uncertainty or tentativeness (``might''). For LLMs, these markers also help infer the appropriate response stance (procedural help, recommendation, speculation, etc.)
    \item \textbf{Speech-Acts and instructions} explicitly name a speaker's intended speech-act (request for assistance, expression of a goal, an instruction to adopt a role/persona, etc.) and suggest the expected interaction outcome. These reflect classic pragmatics theory crucial to conversational studies~\cite{austin1975things,searle1969speech} and the LLM scenarios~\cite{ma-etal-2025-pragmatics}.
    \item \textbf{Use of Pronouns} is fundamentally correlated with human-AI interaction. For instance, addressing the LLM (``you'') and involving oneself (``I'') indicates a conversational setup, and deictic references like ``this'' imply cooperative common ground. These are central to major topics such as whether users perceive AI more as a collaborator than a tool~\cite{schroderqualdata,gaollmtaxonomy}, or the extent of implicit anthropomorphism~\cite{michelle24anth}.
\end{itemize}
For each type, we choose three keywords that (1) distinctly represent specific pragmatic context key to human-AI interaction~\cite{ma-etal-2025-pragmatics}, and (2) is used by $>1\%$ of core users in the first 5 sessions. This creates a total of 13 keywords.\footnote{We later further distinguish the direct addressing of ``you'' and the possessive ``your'', leading to one more sample in the ``Pronouns'' type.}

\begin{table*}[t!]
\caption{Fixed effects of priming ($\beta_1$) and natural long-term adaptation ($\beta_2$), after user-level variance is controlled ($b_u^w$).}
\renewcommand{\arraystretch}{1.05}
%\begin{adjustbox}{width=0.65\textwidth,center}
\begin{tabular}{V{2} c V{2} c V{2} c | c | c V{2} c | c | c V{2}}
\thickhline
\multirow{2}{*}{Type} & \multirow{2}{*}{Keyword} & \multicolumn{3}{cV{2}}{Effect 1: \textbf{Priming}} & \multicolumn{3}{cV{2}}{Effect 2: \textbf{Natural Long-term}} \\ \cline{3-8}
 & & \textbf{$\beta_1^w$} & Sig. & Odds Ratio & \textbf{$\beta_2^w$} & Sig. & Odds Ratio \\ \thickhline
\multirow{3}{*}{Politeness} & please & 4.37 & *** & 79.1 & 0.00062 & ** & $\approx 1$ \\ \cline{2-8} 
 & thank & 3.59 & ** & 36.1 & -0.0058 & N/S & 0.994 \\ \cline{2-8} 
 & hi, / hi\verb*| | & 5.49 & *** & 241.7 & -0.0058 & * & 0.994 \\ \thickhline
\multirow{3}{*}{Modal} & can you & 3.93 & *** & 51.1 & -0.002 & *** & 0.998 \\ \cline{2-8} 
 & should & 4.10 & *** & 60.1 & 0.0023 & * & 1.002 \\ \cline{2-8} 
 & may / might & 2.57 & * & 13.0 & -0.00067 & N/S & $\approx 1$ \\ \thickhline
\multirow{3}{*}{SA \& Inst.} & want & 3.75 & *** & 42.7 & -0.0031 & *** & 0.997 \\ \cline{2-8} 
 & help & 4.69 & *** & 109.0 & -0.0013 & N/S & 0.999 \\ \cline{2-8} 
 & act\verb*| | & 4.28 & *** & 72.3 & -0.0077 & *** & 0.992 \\ \thickhline
\multirow{4}{*}{Pronouns} & you\verb*| | & 2.80 & *** & 16.4 & -0.0008 & *** & 0.999 \\ \cline{2-8} 
 & i\verb*| | & 2.94 & *** & 18.9 & -0.00076 & ** & 0.999 \\ \cline{2-8} 
 & your & 5.15 & *** & 173.0 & -0.016 & *** & 0.984 \\ \cline{2-8} 
 & this & 2.62 & *** & 13.7 & 0.0016 & * & 1.002 \\ \thickhline
 \multirow{2}{*}{(Baseline)} & b & 0.87 & *** & 2.4 & 0.00009 & N/S & $\approx 1$ \\ \cline{2-8} 
 & ,\verb*| | & 1.21 & *** & 3.4 & -0.00075 & *** & 0.999 \\ \thickhline
\end{tabular}
%\end{adjustbox}
\Description{All keywords have a statistically significant fixed-effect of priming, with odds ratio ranging from 13.0 to 241.7. For the effect of natural long-term adaptation, 4 of the keywords do not see a significant effect, and the odds ratios are very close to 1, indicating a rather small impact.}
\label{tab:early_occurrence_regression_result}
\end{table*}

The patterns and results are shown in Table~\ref{tab:early_occurrence_freq_ratio}.
Surprisingly, we see that early interactions are associated with a significantly higher likelihood of being used later on.
In fact, all 13 keywords occur more often by 5.7 to 49.4 times over the rest of a user's life cycle, regardless of duration.

\paragraph{Semantically meaningful priming vs. Arbitrary priming}
To examine whether the trend is relevant to pragmatic implications of language use instead of simply reusing \textit{any} past text, we compare the increase in the keywords against two pseudo-baselines that are not directly associated with any single interpretation: a random letter that is not part of any listed keyword (``b''), and a punctuation plus whitespace (``,\verb*| |'').
We refer to these as pseudo-baselines because any text snippets are inevitably associated with possible keywords (e.g., ``b'' in ``be/based/but/...'') and interpretations. % (e.g., ``, '' can be viewed as at least partially \textit{conversational}).
However, these are sufficiently non-deterministic, not mapping to any single pragmatic scheme or instantly informative keywords.
These baselines are far less impacted by early occurrences than the four types of keywords in question, with $\sim$$2\times$ in the later interactions.

\paragraph{Priming effect vs. Selection Bias}
One potential confounding explanation is selection bias: Some users, for instance, might be simply more polite in general and say ``please'' habitually.
Therefore, they use ``please'' more often in both their first few sessions and later ones for this reason, rather than impacts of early exploration.
To this end, we set up a mixed-effects logistic regression model to examine if there is a significant impact of priming apart from naturally occurring user-level variance.
Our goal is to test the fixed effect $\beta_1$ of $\text{Prime}_u$, the indicator of whether early priming is involved for a user $u$, measured by the presence of $w$ in the first 5 sessions created by $u$.
Separately, a per-user intercept $b^w_u \sim \mathcal{N}(0, {\sigma_w}^2)$ models inherent user-level variance, i.e., the base tendency for $u$ to include keyword $w$ in their language.
We also control and observe the effect of natural change over time as another fixed effect $\beta_2$ of session number $k$.
This helps set aside the alternative dynamics that user behaviors keep evolving, i.e., \textit{long-term adaptations}, contrary to early molding.
Put together, the regression model predicts the binary outcome whether $u$ would use keyword $w$ in their $k$-th turn, $Y^w_{u,k} \sim \text{Bernoulli}(p^w_{u,k})$, with odds denoted as
\begin{center}
    $\text{logit}(p^w_{u,k}) = \beta_0^w + \beta_1^w \cdot \text{Prime}_u + \beta_2^w \cdot k + b_u^w$ \\
\end{center}
Table~\ref{tab:early_occurrence_regression_result} displays the results (significance level and odds ratio) for both effects $\beta_1$ and $\beta_2$ when user-level variance is controlled.
All patterns show a statistically significant fixed effect of early priming. For all four pragmatics-motivated types, keyword patterns see high odds ratios from 13.0 to 241.7. In comparison, the odds of $\beta_2$ is much smaller, and often less or non-significant.
The most notably primed text patterns include greeting (``hi,'', 241.7), addressing the LLM with pronoun ``you'' (173.0), and the instruction to ``help'' (109.0).
%We test other cases (count instead of boolean, window size) in Appendix ...
One example for strong early priming is the auxiliary instruction of ``help''ing (e.g., ``... \textit{help me} create ...'' instead of ``... create ...''): the long-term adaptation yields no significant impact; yet, there is an impressively high effect of early priming (OR$\approx$109).
This verifies the massive impact of earliest interaction not accounted by individual user variance.
Meanwhile, we also note the occasional different order from frequency: for instance, ``thank'' is more significant than ``please'' measured by frequency ($25.3\times$ vs. $11.5\times$), but less so when controlled ($36.1$ vs. $79.1$ OR). This confirms that some patterns are likely more subject to individual variance, i.e., a small group of ``polite'' user habitually thanking, including but not just for LLMs.

\begin{figure}[t!]
    \includegraphics[width=0.3\textwidth]{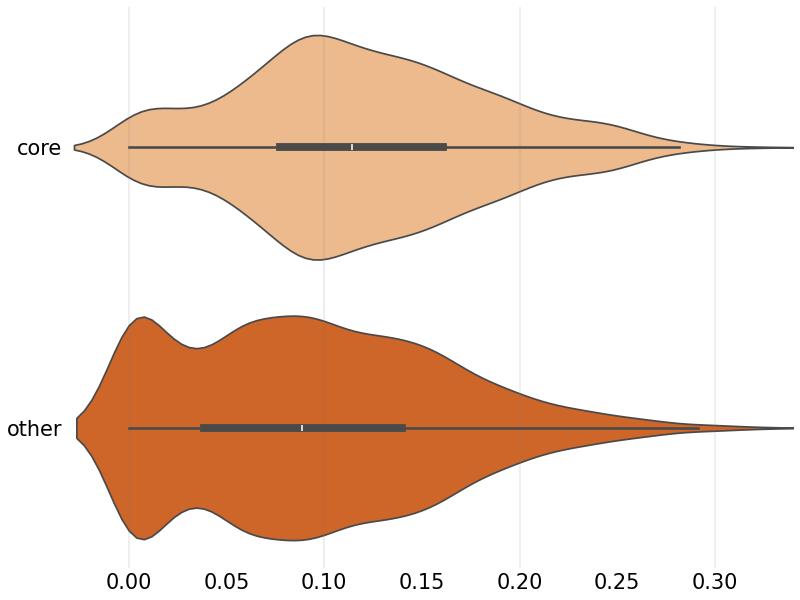}
    \caption{Variety of expressions in the first 5 dialogs, measured by the distances in the first 5 dialogs. The violin plot shows the estimated (KDE) distribution of the average mutual distances of the 5 request/expression components.}
    \label{fig:long-vs-short_avg._distance}
    \Description{Non-core users' distances between expressions are distributed closer to 0 in early interactions.}
\end{figure}

\begin{table}[t!]
\caption{
The effect of diversity of early exploration, under window size $k$ and the corresponding sample size $N$. This can be directly interpreted as a Survival Time Ratio: the ratio of survival time with vs. without the effect ($e^{\beta_1 \bar{d}}$). We display the theoretical full Time Ratio (with $\bar{d}: 0 \rightarrow1$) and contextualize it with two practical units of increase (per 0.1 and 0.01) given the range of cosine distance $d$. For instance, under $k=4$, every 0.01 increase in $\bar{d}$ is expected to extend the user's interaction ``lifespan'' by 5.6\%. The results are consistent and statistically significant, and are validated by Likelihood-Ratio tests.
}
\renewcommand{\arraystretch}{1.25}
\resizebox{\linewidth}{!}{
\begin{tabular}{|c|c|cc|ccc|cc|}
    \hline
    \multirow{2}{*}{$k$} & \multirow{2}{*}{$N$} & \multicolumn{2}{c|}{Main Effect} & \multicolumn{3}{c|}{Survival Time Ratio ($e^{\beta_1 \bar{d}}$)} & \multicolumn{2}{c|}{LR tests} \\ \cline{3-9}
     &  & \multicolumn{1}{c|}{$\beta_1$} & Sig. & \multicolumn{1}{c|}{Full} & \multicolumn{1}{c|}{$\bar{d}\mathrel{+}=0.1$} &  $\bar{d}\mathrel{+}=0.01$  & \multicolumn{1}{c|}{$\chi^2$} & Sig. \\ \hline
    3 & 7955 & \multicolumn{1}{c|}{4.59} & *** & \multicolumn{1}{c|}{98.8} & \multicolumn{1}{c|}{+58.3\%} & +4.7\% & \multicolumn{1}{c|}{266.6} & *** \\ \hline
    4 & 6191 & \multicolumn{1}{c|}{5.45} & *** & \multicolumn{1}{c|}{233.8} & \multicolumn{1}{c|}{+72.5\%}  & +5.6\% & \multicolumn{1}{c|}{246.7} & *** \\ \hline
    5 & 5040 & \multicolumn{1}{c|}{5.70} & *** & \multicolumn{1}{c|}{299.9} & \multicolumn{1}{c|}{+76.9\%}  & +5.9\% & \multicolumn{1}{c|}{196.0} & *** \\ \hline
\end{tabular} %
}
\label{tab:survival_analysis}
\Description{Under all three early exploration window sizes $k=3, 4, 5$, early exploration diversity has a statistically significant effect on long-term retention. For every 0.01 increase in the average distances between the first $k$ sessions of a user, the expected time of use is extended by 4.7 to 5.9\%.}
\end{table}

\subsubsection{Variety in early exploration directly associates with long-term retention.}
The analyses of text pattern reuse have demonstrated the dramatic variances starting from seemingly random early interactions.
We proceed to consider a more fundamental impact besides how users create inputs: does early exploration even relate to the full lifecycles of users, e.g., their tendency to use the system in the long term?
Specifically, is more active early exploration associated with (or even predicts) a higher likelihood to continue using the system?
We inspect the first $k$ sessions of a user: The extent of early expression diversity is calculated as the average cosine distance between the expressions of each pair from one's earliest $k$ sessions.

We start by comparing the earliest expression of the ReCCRE core user group with more sustained activity ($\ge15$ days and $\ge10$ sessions) with other non-core users.
We observe that as early as the first 5 turns, long-term users are sending much more diverse inputs:
As shown in Figure~\ref{fig:long-vs-short_avg._distance}, the variety of expression in the first $n=5$ turns and their comparison between the two user groups differs significantly. Variety is measured as %how many different types total in Figure~\ref{fig:long-vs-short_early_type_num}, and as 
the average distance between all session pairs in Figure~\ref{fig:long-vs-short_avg._distance}, with a statistical difference ($p<<0.001$) visualized in Kernel Density Estimation.
% This higher diversity holds not only for the request components but also for expressions.
% That is, long-term users try more tasks early on, and in the meantime, they are also using more varied language to host their (more diverse) input tasks.

To systematically model the outcome of early exploration, we perform a survival analysis for the impact of more diverse early input styles on user retention. We fit an Accelerated Failure Time (AFT) model with a Weibull distribution~\cite{wei1992accelerated}, a well-established survival analysis framework that directly estimates whether and how much a factor accelerates (or decelerates) one's expected ``life course''. We predict the remaining time of use (dependent variable) given the average cosine distance within the first $k$ sessions (independent variable), $\bar{d}_{u,k}$. Formally, the model is written as
\begin{center}
    $\log(T_u) = \beta_0 + \beta_1\bar{d}_{u,k} + \sigma\epsilon_u$
\end{center}
where $T_u$ represents the expected total remaining days of use for user $u$; $\beta_0$ represents the baseline intercept; $\beta_1$ is the coefficient for the main effect; and $\sigma$ and $\epsilon$ represent the standard Weibull scale parameter and error term.\footnote{the analyses are properly landmarked, i.e., conditioning on users with $>k$ sessions, which correspond to different full sample size $N$ in Table~\ref{tab:survival_analysis}. Here we consider days of use over session numbers for more robust measurement of \textit{long-term retention/survival}, as users may create multiple sessions during the first contact but no longer interact with the model later.}

Table~\ref{tab:survival_analysis} displays the result of the AFT model.
As indicated by the main effect $\beta_1$, higher diversity of early exploration substantially extends the expected platform survival compared with the base survival time, and this impact is consistent across choices of early exploration window size ($k=3, 4, 5$). In fact, for each 0.01 increase in the average cosine distances in a user's expressions early on ($\bar{d}_{u,k}$), the total days of use is projected to increase by 4-6\% (or 50-80\% increase per $\bar{d}_{u,k} \mathrel{+}= 0.1$)! This confirms that active early exploration of different inputs has a significant effect on the possibility of long-term retention, highlighting the need to understand and scaffold these early exploration phases before molding.

\subsection{Interventions and Incentives: What adds to the observations?}
\label{subsec:Experiments_interventions}

Analyses so far regarding user-crafted expressions seem to suggest a unidirectional forming and stabilization.
In this final part of our study, we ask: are there other factors that induce a user to organize and adjust their expressions, apart from the molding over time? Specifically, what might ``intervene'' with the unidirectional molding and motivate a user to further (re-)explore a familiar LLM service?
We consider user input variations under two highly relevant factors in human-AI interaction:
\begin{itemize}
    \item{\textbf{Task-stratification}}: A user can interact with LLMs for numerous purposes, but they may or may not specifically adapt their language to specific tasks (e.g. using casual and close language for companionship, or succinct and tool-like instructions for coding). We explore task-level effects: First, do users really organize their input language based on the request content types? Further, comparing \textit{across different tasks}, how do such impacts vary? What tasks would lead users to apply a unique set of language patterns, and what tasks will be applied the go-to generic interaction modes?
    \item{\textbf{Switching to a different model}}: With the base model constantly updating to newer LLMs, we ask: When users are equipped with prior experience and move on to a different model, will their interaction models resemble an initial exploration process or the stabilized late-stage patterns?
\end{itemize}
The two dimensions also represent a contrast of active vs. passive dynamics on top of molding over time: Organizing input language by task type reflects how a user might manage their inputs and apply local adaptations, whereas an update of the base model probes how an external perturbation interacts with the momentum.

\subsubsection{Users organize input expressions by tasks, but the practical effects vary drastically by task type.}
\label{subsubsec:task_stratification}

To quantify task-specific divergences, we first classify users' initial prompts into several key Human-AI Interaction domains.
In practice, we consider the following types that represent the most common LLM usages and have received extensive interest in the relevant literature:
\begin{itemize}
    \item \textbf{Creative Writing}: The user prompts the system to generate original narrative or expressive content.~\cite{LeeCoAuthor,YuanWordCraft,MirowskiScreenplays}
    \item \textbf{Coding}: The request centers on software development, including writing new code, debugging existing logic, or explaining relevant documentation or syntax.~\cite{WeberProductivity,sarkar2022like,BarkeCopilot,PratherWeird}
    \item \textbf{Editing and Language Transformation}: The user asks for constrained text editing based on given inputs, e.g. to summarize, translate, polish writing, etc.~\cite{cheng-etal-2022-mapping,DangSummaries,KimParaphrasing,zhang2025navfog}
    \item \textbf{Ideation and Planning}: Strategic thinking or open-ended brainstorming aimed at generating lists of ideas, outlines, plans, etc.~\cite{hao-etal-2025-large,ShaerBrainWriting,li2025review}
    \item \textbf{Problem Solving}: Reasoning-intensive interactions such as logical deduction, mathematical calculation, or solutions to objective problems.~\cite{WangTaskSupportive,doi:10.1073/pnas.2318124121,feng-etal-2024-large}
    \item \textbf{Social and Emotional Support}: Interactions where the user seeks empathy, personal advice, companionship, or other similar bonds with the LLM.~\cite{PanSocialSupport,ZhengEmotionalSupport,ma2024understanding,skjuve2021my,sharma2023human}
\end{itemize}
Our goal is to compare the use of \textit{expressions} across this set of \textit{tasks}.
We therefore make use of the Request components ([REQ] in Fig.~\ref{fig:user_example}) to see how the expressions vary under the request types in question.
We use GPT-5-mini to classify a session as one of the six tasks or an additional \textbf{Other} category based \textit{only} on a concatenation of the Request segments, and generate a vector embedding based \textit{only} on the Expression.

Our analysis compares the expressions for different tasks based on their \textit{centroids}.
For each task $T$, we locate a centroid of the expression embeddings, averaging all inputs labelled as $T$:
$$\mathbf{C}^u_{T} = \frac{1}{|C^u_T|} \sum_{c \in C^u_T} c^u_i$$
This is in contrast with a general ``leave-one-out'' centroid of all other sessions not of this task type:
$$\mathbf{C}^u_{\setminus T} = \frac{1}{|C^u - C^u_T|} \sum_{c \notin C^u_T} c^u_i$$
We define the difference between these two centroids, $d(\mathbf{C}^u_{T}, \mathbf{C}^u_{\setminus T})$, as the \textit{displacement} for each task.
If users organize their inputs to format similar tasks in certain ways and deliver different tasks distinctly, each task shall incur a significant displacement, pushing the expressions to be different from the others;
Conversely, if task-stratification is not a major factor, the subset of a specific task type would resemble random sampling. It is also possible that users organize their expression uniquely for \textit{some} tasks (high displacements), but simply apply a more default set of inputs for other tasks (low displacements).
To contextualize the significance of these distances, we create randomized ``tasks'' by sampling 3 to 5 sessions from each user. If users format their inputs for a task in meaningful ways, the centroid displacement should be significantly higher than this random baseline.

\begin{figure}[t!]
    \centering
    \includegraphics[width=0.4\textwidth]{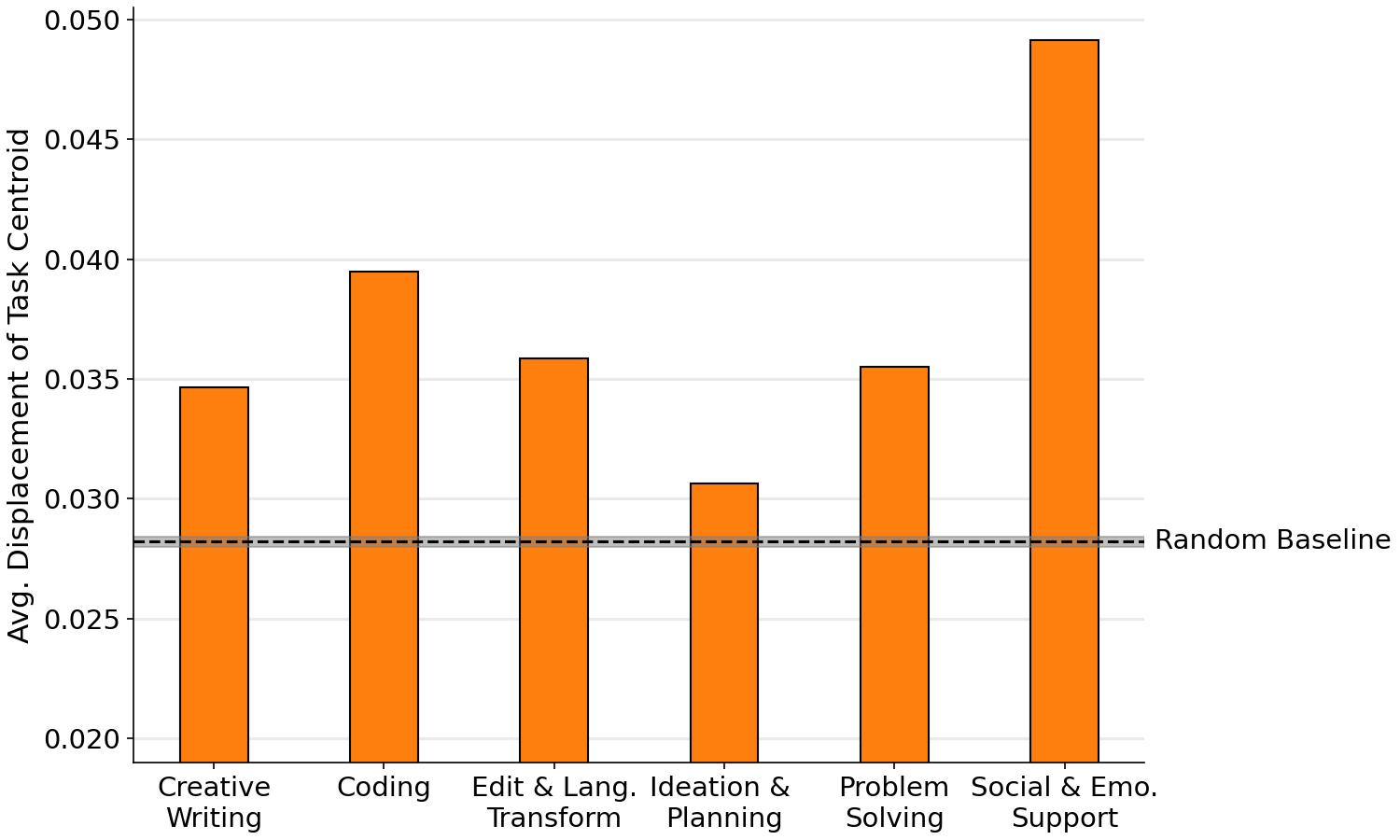}
    \caption{The displacement of task centroids for all six task types. For comparison, we construct a random baseline by sampling 3-5 sessions for each eligible user as a pseudo ``task'', similarly calculate their centroid and displacement, and average across users. The mean of 30 such random trials, aggregated as the random baseline, is shown with a dashed line; a narrow gray region of shades shows 95\% Confidence Invervals of randomization: [0.0279, 0.0284].
    }
    \Description{The displacement for each task are: Creative Writing -- 0.0346; Coding -- 0.0395; Editing and Language Transformation -- 0.0359; Ideation and Planning -- 0.0307； Problem Solving -- 0.0355; Social and Emotional Support -- 0.0492. Random baseline is 0.0281.}
    \label{fig:task_stratification}
\end{figure}

Figure~\ref{fig:task_stratification} shows the average centroid displacement across all eligible users for each task, along with the averaged random baseline and 95\% confidence intervals based on 30 trials. %(footnote?) To ensure robustness and eliminate noise, we apply ... : a user was only analyzed for a target task T if they possessed at least three sessions of T and at least three sessions of non-T tasks.
We first notice that all tasks consistently produce higher displacements than the random baseline, indicating that task types are indeed relevant to user strategies for formatting and organizing their dialog sessions.
Among the tasks in discussion, Ideation and Planning sees the least difference from the random baseline level, only about 10\% more than random. The tasks of Creative Writing, Coding, Editing and Language Transformation, and Problem Solving produce a comparable displacement around 15-20\% higher than Ideation and Planning.
A very interesting finding is the most distinct expressions for Social and Emotional Support: the task displacement is nearly twice the random baseline, and over 25\% more than the highest of other task types (Coding). This demonstrates that, when seeking emotional support and socializing with the chatbot, users consistently lean towards employing specific expressions that are significantly distinct from their overall strategies.

\begin{figure}[t!]
    \centering
    \includegraphics[width=0.35\textwidth]{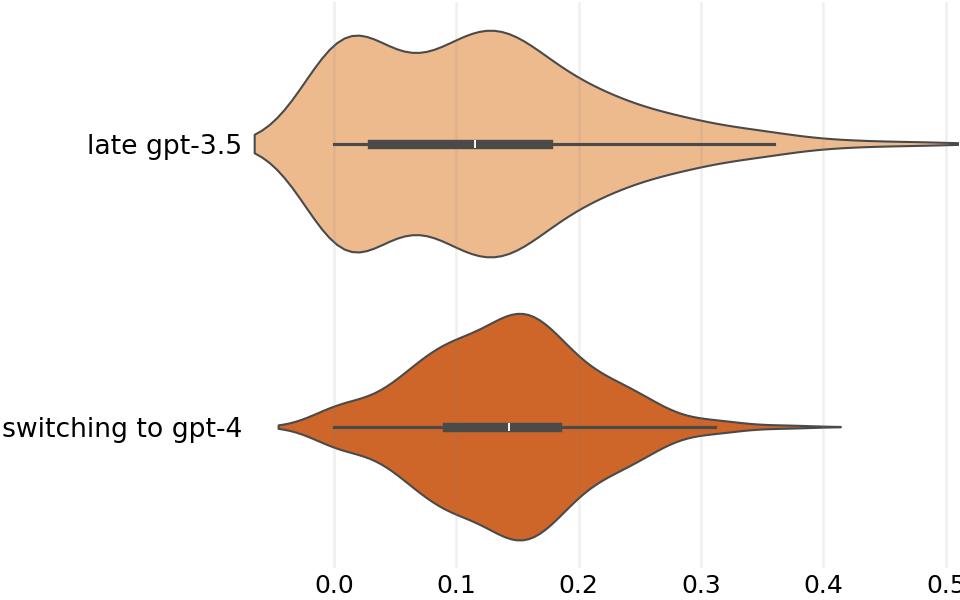}
    \caption{Comparing the expression variety of 246 users as they switch from GPT-3.5 to GPT-4. Significant difference ($p<0.0001$) is found in the average distances of a user's expressions within their last few sessions with GPT-3.5 (top) vs. with their first few sessions switching to GPT-4 (bottom). }
    \label{fig:model_switch}
    \Description{Distances at the last few GPT-3.5 interactions are closer to 0, while that between GPT-3.5 and the first few GPT-4 are centered between 0.1 to 0.2.}
\end{figure}

\subsubsection{Updating to a new model may re-initiate an exploration process}
\label{subsubsec:Experiments_model_change_yes_expression_diff}
WildChat~\cite{zhaowildchat} is operated via self-serve playgrounds on HuggingFace over two GPT model options but asynchronously: GPT-3.5, available throughout the scope of the project; and GPT-4, fully set up starting six months later in October 2023.
This provides a naturally occurring ``intervention'' determined by users: they have built up their experience and mental model with one model and may choose to try out a new one when available.
As users switch to an unseen model, two major factors are antagonistic: the knowledge and experience from the previous model, and the dynamics of initial exploration for a ``new'' model.
We evaluate which factor is dominant in a real-world model update scenario.

We look for users in the dataset that have interacted with both models to create more than 3 sessions.
This results in 246 users carrying their experience with GPT-3.5 and chose to explore the new one based on GPT-4 when it became available.
We use the cosine distance measure to evaluate the average mutual distances of expression in two settings: (1) within the last $n\le5$ sessions with GPT-3.5; and (2) between the last $n$ sessions of GPT-3.5 and the first $n$ sessions of GPT-4.
From the Recency discussions, the former tends to converge; thus, we compare the latter with this baseline level at the point of switching.
We confirm that there is a significant difference ($p=2.1\times10^{-6}$).
Figure~\ref{fig:model_switch} shows the corresponding KDE comparison. We see that the last GPT-3.5 interactions are clustered towards a lower, near-zero difference, consistent with the earlier findings; however, as users switch to GPT-4, the expressions show a significant difference centered between 0.1 to 0.2 instead, comparable to the starting stage of a typical user's initial explorations.

\section{Discussion}

Our large-scale analysis of in-the-wild user-LLM interaction reveals a powerful and rapid process of behavioral formation, where users quickly settle on a small, stable repertoire of interaction (expressions).
This process appears highly individual and path-dependent, and may significantly impact a user's long-term behavior.
%These findings give rise to an \textbf{agency paradox}: the unbounded flexibility of the LLM input space, rather than encouraging exploration, could in fact overwhelm users and end up instead causing a narrowing of behavior patterns and willingness to explore.
These findings provide immediate implications for system design, revisit core assumptions about user adaptation, and raise questions about the future of human-AI interaction and AI literacy.

\subsection{Designing with, not against, the momentum of molding}
\label{subsec:Discussion_design_with_early_interaction}
The most immediate takeaway from our work is that a user's first few interactions are a critical design surface.
With expression patterns crystallizing within 3--5 sessions and persisting 5--50× more frequently in the real world, the window for intervention is remarkably narrow.
Developers cannot afford to ``wait'' for users to organically discover good interaction strategies. Moreover, we observe that long-term activity was distinguished by more varied exploration in their initial sessions.
This calls for a fundamental shift towards more involved design that properly guide interactions.
Practically, we suggest the following factors for future systems:

\subsubsection{Scaffolding interactional and diversified onboarding}
Instead of a passive start, LLM-based dialog systems could take an active role to tutorialize initial interactions, guiding new users through qualitatively different approaches.
This could involve dynamically suggesting alternative phrasings, or presenting a diverse gallery of sample conversations that showcase different interactional modes.
We are in fact not new to this fashion, as various commercial models are already including a few ``try this:''; however, this should be extended from ``guessing what a user might do'' toward showing \textit{different ways to do it}: for instance, whether to use direct instruction or collaborative brainstorming, hints to assign roles when possible, instructions on whether and how to stack different components, etc.
While the goal is not to prescribe a ``correct'' usage, these would seed the user's initial repertoire with more variety, and thus prevent any single pattern from dominating early experience, increase the chance and duration for a user to explore patterns suited for future tasks, and help them build up more complete mental models.

\subsubsection{Strategic friction, ``soft'' intervention, and visualized interactions}
%For existing users who already have established patterns of use, there are still potential interventions.
We discussed that users are not immutable; 
System design can actively seek these incentives for exploration and foster awareness of the gains of such attempts, e.g., creating lower-friction ``soft interventions'' to help users break out of local maxima.
For example, if a system detects a user has used the same indirect phrasing for consecutive, diverse tasks, it might subtly intervene: ``For more complex tasks, you may get better results with more direct, structured instructions. Would you like to see an example?''
This respects user agency while gently hinting at the system's potential.
Hence, the system can proactively create meaningful micro-environment changes, breaking the agency paradox with timely feedback, and thus transform the overwhelming free-form exploration into a collaborative improvement process.
Further, users do not have cross-session perspectives of their explorations, and they cannot optimize what they haven't seen.
Hence, handy support can be provided to help understand the molding process in their own journey, such as a dashboard that visualizes interaction diversity, pattern evolution, and comparative effectiveness~\cite{zhang2026wordswidgetscontrollablellm}.
This could transform unconscious habituation into conscious choice and even a fun space to play with, and motivate continued experimentation.
%Combining with the active nudging, a successful design could change the crucial lack of feedback on such relationships as between expression choices and task outcomes (Recall our findings that currently the expression patterns don't align with task semantics.)
%Making these connections visible and showing how different phrasings yield different results could 

\subsubsection{On the incorporation of complicated dynamics}
Meanwhile, our analysis have also covered the multifaceted nature of user behavior formation.
First, the clear displacements in task-stratified centroids suggest that the molding, as well as the time dimension in general, is not the full story for an unconstrained input space offering significant freedom but little guidance.
In fact, it further raises fundamental questions between a deliberate strategy and a compromising reaction:
Are users coding their expressions into expert schemas, or are they ``trapped'' by the lack of explicit UI affordances and can only rely on stylistic re-phrasing?
For system designers, our analyses call for \textit{context-aware scaffolding} beyond a single blank prompt box.
This tension is most visible in the Social and Emotional Support category, which can often involve transition from a utilitarian mental model to a ``relational'' one, or to prime the model for an empathetic response.
Future design could support such vulnerable interactions with non-linguistic inputs (e.g., affective sliders or explicit persona toggles) that alleviate the cognitive burden of ``writing oneself into a mood,'' thereby preventing the user from feeling trapped in the void of a pure textual space and acknowledge the sensitivity of the task, and eventually provide sufficient and dedicated task-specific affordances.

Another key question is when \textit{not} to leverage or interfere with the priming.
Reduced exploration is not always undesired or problematic: for instance, when the variation of input styles is unnecessary (or even harmful), or when users purposefully apply task-oriented expressions or routine patterns.
If key factors known to incur re-exploration (such as major model updates) are involved, the system should instead adapt to the habituated style and minimize unnecessary effort.
Or, if users specifically approach LLMs with certain interaction modes, e.g., for role-playing or companionship, the system should acknowledge and accommodate the contexts rather than ``optimizing'' inputs.
For instance, OpenAI's rollout of GPT-5~\cite{openaiIntroducingGPT5} was accompanied by the abrupt termination of the previous GPT-4o. While OpenAI claims better performance and less sychopancy~\cite{openaiIntroducingGPT5,openai2025_sycophancy}, it received major backlashes as users found their usual prompting strategies no longer working as effectively, and some users lost the companionship or intimacy they believed to have established with GPT-4o. In fact, users specifically noted GPT-4o's ``warmer'' and ``effusive'' style in emotional support~\cite{Huckins_2025,Freedman_2025}. Here, the assumed narrative of a ``less sycophantic'' LLM fails when it's the exact reason people use it; developers should take into consideration the apparent frictions users would face given the change of style, and provide necessary information and support. %The 4o series was eventually re-released days after it was terminated.

\subsection{The illusion of an ``actively learning user''}
\label{subsec:Discussion_illusion_active_user}
Further expanding our scope of discussion towards a user's entire lifecycle, we question a prevailing assumption underlying user studies: users constantly seek to adapt toward optimal interaction strategies through continued use.
This ``active learner'' narrative expects users to be motivated, acquiring skills and refining their experience; however, it is compromised in the real world, faced with more casual users and a free-form, all-around space for interaction.
Our experiments from real-world data in fact show an opposite trend: users find and stick with a limited scope of input modes and focus on the task content.
Our data also reaffirms that LLM interfaces cannot be neatly mapped onto a learning curve from novice to expert:
We discussed the absence of convergence across users with no universal ``expert'' pattern to discover. Rather, each user picks up their individual trajectories for usable language patterns.
As the process itself plateaus within a relatively short time, these are inevitably subject to individual differences and to randomness.
We suggest reconceptualizing user-LLM interaction not as a user's learning process, but as a habit to be shaped, and a balance to be maintained (or disrupted).

LLM chatbots are trained to be helpful and universally accommodating, providing \textit{some} plausible response to \textit{any} input.
Observations of limited explorations are in line with the design objective to be maximally error-tolerant:
The remarkable flexibility lowers the barrier to entry, but simultaneously limits the incentive for learning.
If a suboptimal prompt still yields an ``okay'' answer, the user receives no clear feedback signal that further grinding is possible or necessary.
But a conversation simply moving forward doesn't guarantee that it is proving the most successful or useful.
A chat interface leading users to pursue one specific trajectory among the practically infinite dialog space --- this may yield a silent misalignment regarding perceived success (or failure) and actions taken:
Users may remain unaware of the variability of what the model \textit{could have} delivered after their early experience where their use styles are molded, and hence settle into a potentially non-optimal mental set.
A user might, for instance, receive a mediocre summary from a generic prompt and assume, ``This is the best the AI can do,'' rather than, ``I should have asked that differently.''

Does a user’s lack of exploration stem from genuine satisfaction with the result? Or, is the satisfaction simply because the system never gave them a reason or motivation for a possible ``better'' result?
This is only possible with first-hand user reactions via human-in-the-loop data practices, which are lacking in datasets like WildChat. (In fact, WildChat doesn't provide \textit{any} user feedback; this creates systematic limitations, and we discuss them in \S\ref{sec:limitations}.)
While recent LLM services are equipped with in-product feedback functionalities such as (dis)liking a response or user-preference A/B tests~\cite{cooper2025constraining,openai2025_sycophancy,kampf2024_compare_mode},
these simple reactions are still insufficient for modeling user experience:
As they don’t directly map to visible differences, users are not adequately motivated to interact with these features, let alone knowing how to utilize them to adapt or improve their experience.
This resonates with the context of rapid molding: probing success or satisfaction is not trivial; unless the output is exceptionally good/bad, users may well just satisfice~\cite{simon1956rational} to the “okay” outputs they get and stick to habitual styles.

\subsection{The future of Human-AI Interaction research with priors}
\label{subsec:Discussion_future_research_with_priors}
Our findings suggest fundamental reconsideration of how HAI research accounts for user experience and evaluates system design. 
%More work will be needed to understand the deeper causes of our data observations.
% Entrenchment may suggest a fundamental struggle: Users are getting by, but not thriving; and the interface does little to reveal potentials.
% However, the rapid stabilization may also be an accomplished goal of system design: A low learning curve allows users to become functional without notable efforts required to ``learn'' the interface. Thus, users are efficiently offloading the work of ``how to ask'' so they can focus on ``what to ask.''
%We invite future works to provide more user-centered, mixed-methods insights towards modeling \textit{user experience} in face of the agency paradox, and thus better support users in their free-form interactions.
As the public continues to gain more experience with LLM/AI, it will be increasingly inaccurate if a study implicitly assumes participants are blank slates and/or continuously optimizing.
Our findings show that people likely walk in with existing interaction habits often unconsciously formed.
Researchers need to incorporate this ``interactional prior'' as a significant variable, and to build and test with controls for past trajectories. 
We suggest that future research take into consideration further steps beyond current practices:

\paragraph{Acknowledging the baseline in-the-wild experience}
As LLMs become ubiquitous, recruiting truly naive users becomes impossible. Every participant arrives with crystallized patterns that influence their study behavior.
A user habituated to brief, imperative commands can interact differently from one who writes conversational paragraphs, despite being under seeming controlled task instructions.
This could suggest a potential ``contamination'', and future studies should account for these differences, actively work to neutralize them through extensive familiarization periods, and discuss the potential impacts of such prior experience as a confounder. %The common practice of brief tutorials before data collection may be insufficient when combating months of habituation.
%This is notably beneficial for robust design improvements, ensuring that a new element has a truly generalizable reach, instead of overfitting to the random quirks brought in by test users.

\paragraph{Prior experience as trajectory, not category}
Current research typically measures experience through duration (``6 months of use'') or frequency (``daily user''), assuming homogeneous skill development.
Our data reveals that users may have entirely different interaction repertoires based on their initial exploration paths.
Researchers should capture not just exposure but interaction biography --- the specific patterns users have crystallized. 
Pre-study instruments could include warm-up tasks that reveal participants' existing patterns, or request examples of participants' recent LLM interactions. 
This helps to shift from asking ``How experienced are you?'' to ``Show us how you typically interact.''

\paragraph{Exploration trajectories as a design objective}
Further, the findings of our work suggest a new objective for designing with LLM/AI: an ideal system should consider not only the system's current state in implementing requests, but also its capacity to shape beneficial and motivating trajectories.
A system that performs well for experienced users with diverse patterns might fail to cultivate such diversity in new users.
Longitudinal evaluation is essential for not just tracking performance over time but examining whether a system has built an environment that encourages exploration, versus one that reinforces narrow patterns.
Metrics might give a specific focus on user expressions, such as an index for expression diversity or ``half-life'' alongside traditional task completion rates.

\paragraph{Making sense of the researcher's dilemma}
Finally, in a broader view, our findings implicate researchers themselves.
The prompts used in studies, the tasks designed for evaluation, and the interaction patterns modeled in tutorials all reflect researchers' \textit{own} crystallized patterns.
Much like any downstream users, we could be accustomed to detailed, structured prompts, and thus design studies that inadvertently favor users with similar patterns.
We note the need for research teams to build with awareness of this subtlety, look for diverse interaction patterns, and try to put explicit efforts to generate variety in study materials.

\section{Limitations and Future Work} 
\label{sec:limitations}

Our study represents a detailed example of a large-scale, longitudinal analysis of in-the-wild user data, which allows us to model authentic behavioral patterns. Nonetheless, this methodological approach carries inherent limitations rooted in the current form of data resources. We conclude with discussions on these nuances, which we believe can open avenues for future research.

First, while our analysis reveals what patterns emerge and persist, there is no conclusion at causal level. This is the nature of existing unstructured real-world chat logs as data~\cite{zhu2025data} without in-lab level control. We observe users settling into narrow interactional repertoires, but a more well-formed data paradigm is required to distinguish between a conscious choice for efficiency and an unconscious inability to discover better strategies. Future work may employ mixed-method approaches to bridge this gap, e.g., first identify users with strong path-dependent behaviors from large-scale logs, and then engage them in semi-structured interviews or diary studies to uncover the mental models that drive their interaction.

Another intrinsic limitation of current data resources is the absence of explicit feedback.
The interpretation of \textit{task success} is crucial, but complicated in a fully open-ended environment.
The lack of any user intention creates challenges of the \textit{unmeasured bias}:
Exploration trajectories still account only for limited and most ``standard'' actions that get documented.
Real-world user dynamics can be chaotic: For instance, a user might intentionally explore via “spamming”, by creating a few parallel sessions that are highly similar or identical just to see what the system would do.
Or, conversely, one might simply leave at any point and not return with or without a specific reason.
%(And – this same user might unexpectedly come back when their OpenAI subscription expires and this free project occurred to them.) 
These important dynamics may be studied in future work (for instance, when and why users choose to silently quit instead of attempting further or expressing dissatisfaction).

Besides, our work focuses on the initial (first-turn) user input, as it is where the user freely uses expressions to start a session. However, the whole picture of user-LLM interaction relies on full sessions; the latter parts not covered in this work may be interesting in its own way --- for example, the back-and-forth revisions and types of follow-up actions~\cite{mysore-etal-2025-prototypical}.
These would involve carefully designed data collection (e.g., designing UI for real-time feedback; mining patterns of satisfaction in the corpus; etc.), and we encourage future work to explore new paradigms. %It is also important to blend in more specific demographic and situational context in creating these sessions, even when task content variety are accounted for.

Finally, our findings are (still) a snapshot based on the models and interfaces prevalent in 2023-2024. The landscape of generative AI is evolving at a breakneck pace, and interfaces are also becoming more multimodal and agentic. It is an open question how these emerging AI paradigms will influence the patterns of path dependence we observed. Will voice- or image-based interactions create different kinds of behavioral ruts? Can specialized AI agents effectively serve as the ``soft interventions'' to help users break out of local maxima? Future longitudinal studies are crucial to track these evolving dynamics, using the patterns identified in our work as a baseline to understand the future of human-AI co-adaptation. We believe this work lays the foundation for ongoing investigation into how users learn to work with our increasingly intelligent systems. % Our work also suggests a future for ``legible'' AI systems that don't just respond to requests, but also help users reflect on their own interaction processes, and eventually build an ecosystem where interaction between AI and humans is beyond calling task-solving tools.

\section{Conclusion}
This paper explores real-world user-LLM conversations %Aligning with broader research on conversational interfaces, we inspe how users come into such interfaces with
%and explore the wide variety of preconceptions and interaction patterns. 
following new protocols~\cite{zhu2025requestmaking} to separate request content from task-independent expressions. Our work features a brand new data-driven methodology complementing conventional in-lab studies: understanding the commonality and variety of interaction via the careful analysis of naturally occurring user-LLM chat logs at a million scale. The proposed frameworks further allow us to track broader and authentic threads among segments of users over time. 

We highlight how user interactions are rapidly molded in as few as 3-5 sessions.
User behaviors appear ``sticky'': text patterns attempted in this early period are heavily reused in the long run, and user retention is associated with more active and diverse early explorations.
%Yet, analyses show no ``universal optimal'' but highlights varied individual trajectories.
Through mixed-effects models, we confirmed that this priming effect was not solely due to user-level variance.
Further, we consider various relevant factors: Users seem to organize expressions by task types (but vary dramatically for different tasks), and a natural experiment of model updates did incur a ``re-exploration'' pattern among users.
Meanwhile, these observations come with caveats related to WildChat's lack of success metrics and feedback, which limit how far we can draw conclusions.

We hope this paper will act as a call for further investigation into the evolution of prompting behavior among diverse user groups, and raise attention to designing with this early momentum of users. In the immediate future, more in-lab studies and qualitative interpretations of the molding dynamics can provide rich insights.
%to examine how model version changes have affected these trends, how specific user expertise may mediate the patterns we observed, and how design interventions such as onboarding can shape the ``sticky'' phase of chats among others.
For peer researchers, we also call to systematically take into consideration the priors and molded interaction patterns of users, which are crucial to HAI and UX studies but often inadequately modeled.

% \begin{acks}
% \end{acks}

\bibliographystyle{ACM-Reference-Format}
\bibliography{custom}

\clearpage

%TC:ignore
\appendix

\section{Choice of parameter $\delta$ in \S\ref{subsubsec:bounded_expression_types}}
\label{appendix:params}

\begin{figure}[h!]
    \centering
    \includegraphics[width=0.4\textwidth]{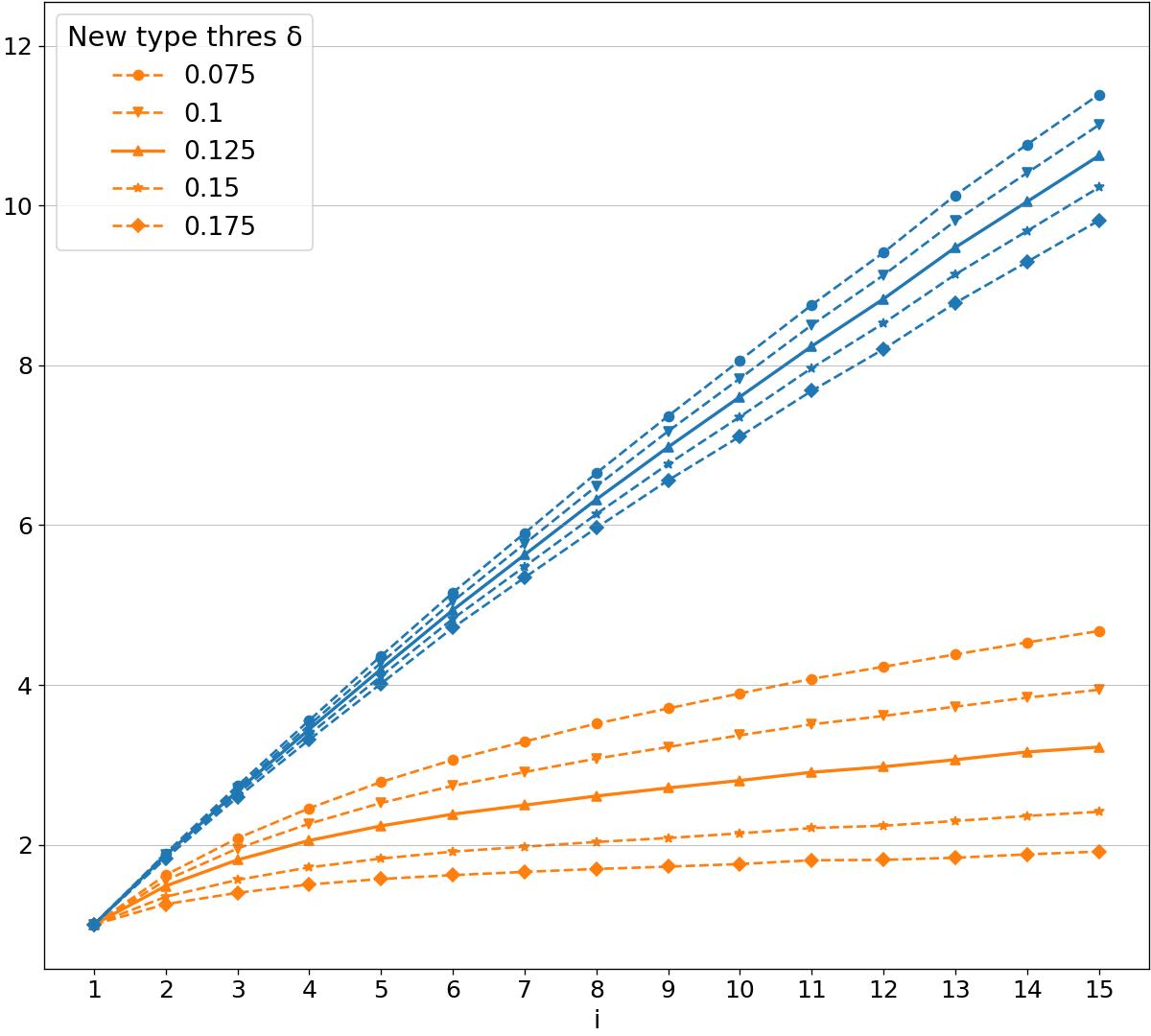}
    \caption{The result of scale comparison of request content vs. expressions (Figure~\ref{subfig:total_type_num_scale-comparison}) under a series of different choices of threshold $\delta$.}
    \label{fig:fit_thres_sensitivity}
    \Description{The figure shows the linear scale of request content types compared with the logarithmic scale of expression types~\ref{fig:total_type_num_and_fit} under a sequence of 5 different choices of $\delta$: 0.075, 0.1, 0.125, 0.15, 0.175. Under all settings, the type number of expressions shows a stabilizing pattern after 5--10 turns, despite variance in the specific number of types (since a smaller and stricter threshold will yield more types), varying from 1.8 for $\delta=0.175$ to around 4.2 for $\delta=0.075$. The types of request content remain relatively stable, with a near-linear trend adding up to around 10--11 types in 15 turns.}
\end{figure}

We show that the findings in \S\ref{subsubsec:bounded_expression_types} are not dependent on any specific choice of the threshold $\delta$ for different ``types'' but are rather insensitive (within reasonable range --- e.g., an unrealistic $\delta=1$ will predict everything to be the same type).
For reference, the average expression distance of the same user under Recency converging after 15+ turns is 0.09--0.1, and that of two random users' first expression is 0.16--0.17 (from Figure~\ref{fig:conv_expression_cosine} and Figure~\ref{fig:dists_across_users_no_convergence}).
We illustrate in Figure~\ref{fig:fit_thres_sensitivity} the effect of $\delta$ with a sequence of 5 values varying from rather small (0.075, lower than late-stage same-user) to excessively large (0.175, higher than random users' expression compared).
As a result, we confirm the robustness of the findings: despite the (expected) overall increase with stricter thresholds, the converging logarithm scale of expression type numbers is visible and consistent under large variations of $\delta$.

%TC:endignore
\end{document}